\documentclass[resetfootnote ]{aastex701}

\usepackage{lmodern}
\usepackage{CJKutf8}
\usepackage{amsmath}
\usepackage{hyperref}
\usepackage{xspace}
\usepackage{caption}
\usepackage{upgreek}
\usepackage{ulem}

\newcommand{\suppmattwo}{Appendix\xspace}
\begin{document}
\title{Time-resolved Polarimetry of the Type Ibn Supernova 2025kzr Suggests a Possible Ejecta–CSM Misalignment}
\shorttitle{Time-resolved Polarimetry of SN 2025kzr}

\author[orcid=0009-0007-3401-7133]{Xudong Wen*\begin{CJK*}{UTF8}{gbsn}
(文旭东)\end{CJK*}}
\affiliation{Department of Physics, Tsinghua University, Qinghua Yuan, Beijing 100084, China}
\email[show]{*wenxudong@mail.tsinghua.edu.cn}  

\author[orcid=0000-0002-6535-8500]{Yi Yang*\begin{CJK*}{UTF8}{gbsn}
(杨轶)\end{CJK*}}
\affiliation{Department of Physics, Tsinghua University, Qinghua Yuan, Beijing 100084, China}
\email[show]{*yi\_yang@mail.tsinghua.edu.cn}  

\author[0000-0001-7092-9374]{Lifan Wang}
\affiliation{George P.\ and Cynthia Woods Mitchell Institute for Fundamental Physics \& Astronomy, Texas A\&M University, 4242 TAMU, College Station, TX 77843, USA}
\email[unshow]{lifan@tamu.edu}

\author[0000-0003-1349-6538]{J. Craig Wheeler}
\affiliation{Department of Astronomy, University of Texas, 2515 Speedway, Stop C1400, Austin, TX 78712-1205, USA}
\email[unshow]{wheel@astro.as.utexas.edu}

\author[0000-0003-3460-0103]{Alexei~V.~Filippenko}
\affiliation{Department of Astronomy, University of California, Berkeley, CA 94720-3411, USA}
\affiliation{Hagler Institute for Advanced Study, Texas A\&M University, 3572 TAMU, College Station, TX 77843, USA}
\email[unshow]{afilippenko@berkeley.edu}

\author[0000-0003-0733-7215]{Justyn~R.~Maund}
\affiliation{Department of Physics, Royal Holloway, University of London, Egham Hill, Egham, TW20 0EX, UK}
\email[unshow]{Justyn.Maund@rhul.ac.uk}

\author[0000-0002-8255-5127]{Mattia Bulla}
\affiliation{Department of Physics and Earth Science, University of Ferrara, via Saragat 1, I-44122 Ferrara, Italy}
\affiliation{INFN, Sezione di Ferrara, via Saragat 1, I-44122 Ferrara, Italy}
\affiliation{INAF, Osservatorio Astronomico d'Abruzzo, via Mentore Maggini snc, 64100 Teramo, Italy}
\email[unshow]{mattia.bulla@unife.it}

\author[0000-0003-2449-1329]{Simon de Wet}
\affiliation{DTU Space, Technical University of Denmark, Elektrovej 327–328, 2800 Lyngby, Denmark}
\email[unshow]{sndwe@dtu.dk}

\author[0000-0002-5221-7557]{Chris Ashall}
\affiliation{Institute for Astronomy, University of Hawai'i at Manoa, 2680 Woodlawn Dr., Hawai'i, HI 96822, USA}
\email[unshow]{cashall@hawaii.edu}

\author[0000-0001-6797-1889]{Steve Schulze}
\affiliation{Department of Particle Physics and Astrophysics, Weizmann Institute of Science, Rehovot, Israel}
\email[unshow]{steve.schulze@weizmann.ac.il}

\begin{abstract}

Type Ibn supernovae (SNe) are rare interacting transients. The dense, helium-rich circumstellar material (CSM) surrounding these events provides insight into the final stages of massive-star evolution. The geometries of the ejecta and CSM provide complementary constraints on the explosion and pre-SN mass loss, yet their temporal evolution has remained largely unexplored by spectropolarimetry. Here we present the first spectropolarimetric time series spanning the interaction-to-ejecta transition of the Type Ibn SN\,2025kzr. 
Three epochs of Very Large Telescope observations, combined with two epochs of published $V$-band imaging polarimetry, span 5--34 days after explosion.
Continuum polarization is low at the earliest epoch, then rises to $\sim$0.6\% by day\,11 as the SN ejecta sweep up the dense CSM. 
This rise, together with a rotating polarization position angle, indicates a significant change in the geometry of the emitting region. 
The polarization subsequently falls below our detection limit as the photosphere recedes into the ejecta.
Radiation-hydrodynamic and Monte Carlo polarization models illustrate that this evolution can be reproduced by an oblate CSM and prolate ejecta, with the symmetry axis of the cold dense shell tilted relative to that of the CSM. 
Binary mass transfer can produce an equatorially enhanced CSM, although the polarization data alone cannot identify the progenitor's formation channel.
Similar polarization signatures in other supernovae suggest that misaligned ejecta and CSM may be common, making time-resolved spectropolarimetry a powerful probe of the link between pre-explosion mass loss and explosion geometry.

\end{abstract}

\keywords{\uat{Core-collapse supernovae}{304} --- \uat{Circumstellar matter}{241} --- \uat{Spectropolarimetry}{1973} --- \uat{Supernovae}{1668} --- \uat{Polarimetry}{1278}}

\section{Introduction}~\label{sec:intro}

Type Ibn supernovae (SNe Ibn) are a rare class of interacting core-collapse supernovae (CCSNe), constituting only $\sim1$\% of the volumetric rate of all CCSNe in the local universe~\citep{2020ApJ...904...35P, 2022ApJ...927...25M,Srivastav2026}. 
They are characterized by relatively narrow He\,{\sc i} emission lines, the full width at half maximum (FWHM) velocities of $\sim$1000\,km\,s$^{-1}$, with little or no hydrogen, indicating strong interaction between the SN ejecta and dense, helium-rich circumstellar material (CSM; see, e.g.,~\citealp{2000AJ....119.2303M, 2007Natur.447..829P}, and reviews by~\citealp{2017hsn..book..195G, 2017hsn..book..403S}).
SNe Ibn show broadly similar post-peak light curves but diverse early-time spectra, which may reflect differences in their progenitor systems or CSM \citep{2017ApJ...836..158H}.
%Their progenitors are thought to be stripped-envelope massive stars that underwent substantial mass loss shortly before core collapse.
Their progenitors remain uncertain, but may include stripped-envelope massive stars that underwent substantial mass loss shortly before core collapse, as well as binary systems in which mass transfer produced dense, H-poor circumstellar material~\citep{2007ApJ...657L.105F,2007Natur.447..829P,2022ApJ...927...25M,2025A&A...696A.103E}.

Proposed mechanisms driving this pre-explosion mass loss include steady stellar winds~\citep{Quataert2016}, pulsational pair instability~\citep{Yoshida2016, Woosley2017}, episodes of eruptive mass ejections triggered by gravity waves or shell burning~\citep{2012MNRAS.423L..92Q, 2014ApJ...780...96S, Woosley2015, Fuller2017, Fuller2018}, and binary interaction~\citep{Soker2013, Smith2014, Smith_Arnett_2014, Yoon2017, Metzger2022, 2022A&A...658A.130D,Tsuna2024,2025A&A...696A.103E}.  
These mechanisms can produce distinct three-dimensional (3D) CSM geometries, making the CSM structure a powerful probe of the progenitor's final evolutionary stages.

Spectropolarimetric observations of interacting SNe, such as Type IIn SNe and SN~Ia-CSM/SN1997cy-like events, indicate that the CSM is often significantly aspherical \citep{2011A&A...527L...6P,2023MNRAS.519.1618Y,2024MNRAS.527.6090M,2024MNRAS.529.1104B}; however, the geometry of the helium-rich CSM in classical Type Ibn SN remains poorly constrained by multiepoch spectropolarimetry. Measuring this geometry can help distinguish among the proposed mass-loss channels, while its temporal evolution can trace the changing contributions of the CSM, the interaction region, and the ejecta, thereby linking pre-explosion mass loss to explosion asymmetry.

Photometric and spectroscopic studies of SNe Ibn have placed important constraints on the radial density profile, total mass, and characteristic velocity of the helium-rich CSM~\citep{2007Natur.447..829P, 2008MNRAS.389..113P, 2017ApJ...836..158H, 2022ApJ...926..125P, 2022ApJ...927...25M}. 
Light-curve modeling, in particular, indicates that the rapid post-peak decline characteristic of SNe Ibn is consistent with a steep radial density profile, $\rho_{\rm CSM} \propto r^{-3}$, implying a mass-loss rate that increased toward core collapse \citep{2022ApJ...927...25M}. Similarly, the overall photometric homogeneity of the SN~Ibn subclass has been interpreted as evidence for a spatially confined helium shell embedded in a less dense, extended wind \citep{2017ApJ...836..158H}.
These one-dimensional model fittings to the light curves, however, are insensitive to the angular structure of the CSM and to its orientation relative to the ejecta. 
Spectroscopic line-profile asymmetries in some SNe Ibn, such as the redshifted nebular He{\sc i} lines of SN\,2015G, have provided indirect hints of geometric asymmetry~\citep{2017MNRAS.471.4381S}, but they cannot uniquely constrain the 3D structure of the emitting region.
It therefore remains unclear whether the helium-rich CSM is distributed spherically, concentrated toward an equatorial plane, clumpy, or arranged in a more complex asymmetric morphology. This geometric degeneracy limits our ability to distinguish among the proposed mass-loss scenarios and to connect the progenitor’s final mass-loss history to the explosion mechanism. 

Spectropolarimetry provides a powerful method for studying the multidimensional geometry of unresolved supernova ejecta and and their surrounding media.
While the source cannot be spatially resolved, a net polarization signal survives only if the scattering geometry deviates from spherical symmetry~\citep{2008ARA&A..46..433W}. 
Continuum polarization arises from electron scattering in an aspherical photosphere, while line polarization across spectral lines traces the distribution of the absorbing and scattering material in the ejecta and CSM. 
This technique has revealed pronounced asymmetries in SNe of nearly all types, yet comparable datasets for Type Ibn events remain extremely scarce. 
Single-epoch spectropolarimetry of SN~2015G revealed substantial continuum polarization (\(P\approx2.7\%\)) \(\sim5\) days after discovery \citep{2017MNRAS.471.4381S}, whereas host-galaxy interstellar polarization (ISP) dominated the near-peak observations of SN~2015U~\citep{2016MNRAS.461.3057S}, precluding firm constraints on its intrinsic signal. 
Broadband measurements of SNe\,2023emq and 2023tsz added further constraints, but both sampled only limited phases~\citep{Pursiainen2023, Warwick2025}. 
Multiepoch polarimetry of the transitional Ibn/Icn SN~2024abvb revealed an evolving polarization signal, declining from \(\sim1\%\) near peak to \(\lesssim0.5\%\) after a week before rising to \(\sim4\%\) by \(\sim30\) days, accompanied by a \(\sim50^\circ\) position angle (PA) rotation, consistent with a changing, asymmetric scattering geometry \citep{2026A&A...710A..35A}.
This points to a time-variable, wavelength-dependent scattering or obscuration component and highlights the potential of time-resolved spectropolarimetry to probe structured CSM.
Nevertheless, how the geometry of the ejecta and CSM relate and evolve across the transition from interaction-dominated to ejecta-dominated emission remains poorly constrained for SNe Ibn.

In this {\it Letter}, we present high-quality, multiepoch spectropolarimetric observations of the Type Ibn SN\,2025kzr, spanning the transition from CSM-interaction-dominated to ejecta-dominated emission. Both the continuum polarization degree and the PA evolve significantly over this period. 
We find that these changes are consistent with a geometry in which the symmetry axes of the ejecta and CSM are misaligned by approximately \(33^\circ\). 
This result parallels the time-dependent polarization observed in the Type II SNe\,2023ixf and 2024ggi; both are interpreted as evidence for misaligned geometries of the ejecta and surrounding CSM, suggesting that such asphericity 
%may be ubiquitous among SN subclasses of Type II/IIP
may be more common than previously recognized among SN subclasses of Type II/IIP~\citep{2023ApJ...955L..37V, 2026ApJ..1000...18V, Yang2025}. 
If such misalignments are confirmed in a larger sample, they would provide a new means of quantifying how common multidimensional structures are in core-collapse explosions and their surrounding environments. 

This paper is organized as follows. Observations, data reduction, and the estimation of the interstellar polarization are discussed in Section~\ref{sec:specpol}. Section~\ref{sec:analysis} describes the time evolution of the continuum and line polarization of SN\,2025kzr. Modeling of the temporal evolution of the continuum polarization is carried out in Section~\ref{sec:model}. A discussion and brief summary are given in Section~\ref{sec:discussion}.

\section{Imaging and Spectropolarimetry of SN~2025kzr}~\label{sec:specpol}
We obtained three epochs of spectropolarimetry of SN\,2025kzr using the FOcal Reducer and low-dispersion Spectrograph 2 (FORS2;~\citealp{1998Msngr..94....1A}) on Unit Telescope 1 (UT1, Antu) of the ESO Very Large Telescope (VLT), as listed in Table~\ref{Table_pol}. All observations were carried out in the Polarimetric Multi-Object Spectroscopy (PMOS) mode, with the 300V grism in place and coupled to a 1$''$-wide slit. 
This configuration provides a wavelength coverage of $\sim 3500$--9200\,\AA\ and a spectral resolution of $R\approx 440$ at the central wavelength of 5849\,\AA, corresponding to a resolution element of $\sim 13.3$\,\AA.
To maximize the blue wavelength coverage, we did not use the GG435 order-sorting filter, which has a cutoff wavelength of $\sim$4350\,\AA, since the contamination by second-order light has a negligible effect on the extracted polarization signal (see Appendix of \citealp{2010A&A...510A.108P}). 

Details of the FORS2/PMOS data reduction and the derivation of the Stokes parameters can be found in the FORS2 Spectropolarimetry Cookbook and Reflex Tutorial\footnote{\url{ftp://ftp.eso.org/pub/dfs/pipelines/instruments/fors/fors-pmos-reflex-tutorial-1.3.pdf}}, \citet{2017MNRAS.464.4146C}, and in Appendix~A of \citet{2020ApJ...902...46Y}, following the procedures described by~\citet{2006PASP..118..146P} and~\citet{2007MNRAS.381..201M}. 
Because the sensitivity of FORS2 decreases rapidly below $\sim 4000$\,\AA, leading to large polarization uncertainties at the blue end of the spectrum, we do not attempt to characterize polarization features at shorter wavelengths.
Table~\ref{Table_pol} presents the log of VLT spectropolarimetry for SN\,2025kzr, along with the continuum polarization measured at each epoch (see Section~\ref{subsec:Continuum Polarization}).

The raw ordinary- and extraordinary-beam spectra were preprocessed following standard routines (bias subtraction and flat-field correction), and the total-flux spectra were then extracted with IRAF\footnote{{IRAF} is distributed by the National Optical Astronomy Observatories, which are operated by the Association of Universities for Research in Astronomy, Inc., under cooperative agreement with the U.S. National Science Foundation.}~\citep{1986SPIE..627..733T, 1993ASPC...52..173T}. 
Cosmic rays were removed with the Laplacian edge-detection algorithm L.A.Cosmic~\citep{2001PASP..113.1420V}. The wavelength calibration of each spectral trace reached a typical root-mean-square (RMS) accuracy of $\lesssim$0.25\,\AA. The subsequent polarimetric analysis, including the calculation of the Stokes $Q$ and $U$ parameters and polarization spectra~\citep{2006PASP..118..146P, 2007MNRAS.381..201M}, the correction for instrumental polarization~\citep{2007ASPC..364..503F, 2014A&A...561A..82S, 2017MNRAS.464.4146C}, and the bias correction needed to recover the true polarization level~\citep{1985A&A...142..100S, 1997ApJ...476L..27W}, is described in detail by \citet{2017MNRAS.464.4146C}, \citet{2020ApJ...902...46Y}, and \citet{2023MNRAS.519.1618Y}.

%Figure~\ref{fig:1} presents the evolution of the flux and spectropolarimetric properties of SN~2025kzr from 8 to 33.9\,d.
%The Stokes $Q$ and $U$ spectra, as well as the debiased polarization degree, are shown after subtraction of the adopted ISP, whereas the polarization position angle is displayed prior to the ISP correction. The polarization spectra are rebinned to 50\,\AA\ for clarity.

\begin{figure*}
    \centering
    \includegraphics[width=0.9\textwidth]{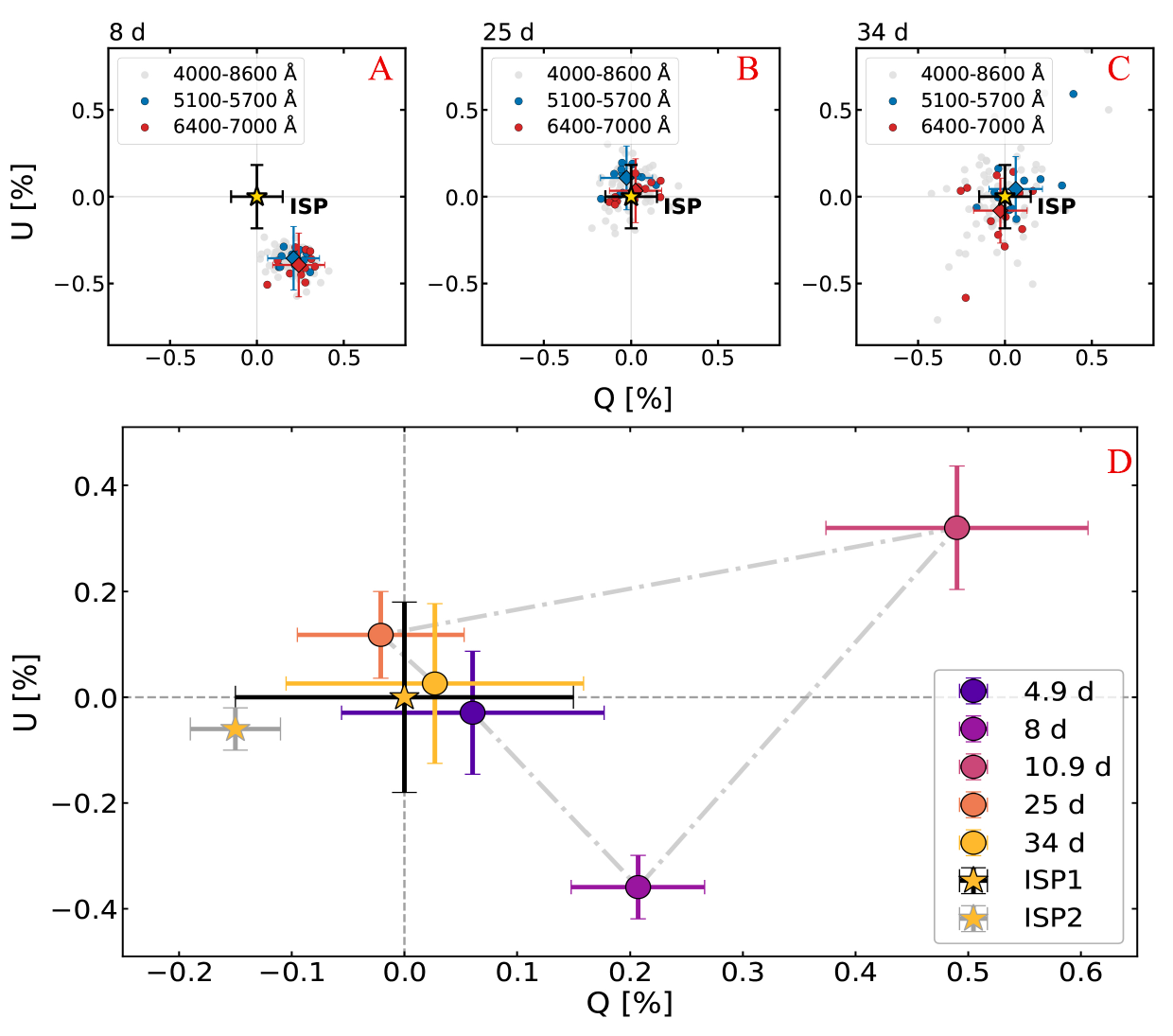}
    \caption{
    Polarization of SN\,2025kzr displayed on the Stokes $Q-U$ plane. Panels A–C show the measurements at 8.0, 25.0, and 34.0\,d, respectively, where gray, blue, and red points denote the 4000--8600, 5100--5700, and 6400--7000\,\AA\ wavelength bands. Panel~D shows the time evolution of the continuum polarization in V band, with colors identifying the epochs as indicated in the legend. Star symbols mark the two adopted ISP estimates, ISP1 and ISP2. Error bars indicate the measurement uncertainties, and the gray dash-dotted lines connect successive measurements.}
    \label{fig:1}
\end{figure*}

\setlength{\tabcolsep}{3.5pt}
\begin{table*}[t]
\begin{center}
\caption{VLT Spectropolarimetic Observations of SN\,2025kzr.}~\label{Table_pol}
\begin{normalsize}
\begin{tabular}{c|c|c|c|c}
\hline
\hline
Epoch  &  MJD-OBS / Phase$^a$  &  Exp. Time$^b$  &  $Q^{c}_{\rm Cont}$ & $U^{c}_{\rm Cont}$  \\
 \#   &  (day)      &  (s)            &  [\%]           &  [\%]     \\
\hline
1      & 60824.022 / 8.0 &  $2\times4\times150$  &     0.242$\pm$0.076 & $-$0.394$\pm$0.067   \\
2      & 60840.971 / 25.0 &  $2\times4\times360$  &   $0.026\pm$0.099 & 0.033$\pm$0.058   \\
3      & 60849.980 / 34.0 &  $2\times4\times360$  &   $-$0.026$\pm$0.126 & $-$0.081$\pm$0.188  \\
\hline
\end{tabular}\\
{$^a$Relative to the estimated explosion epoch of UTC 2025-05-21 (MJD $60816.0\pm0.1$; \citealt{deWet2026}).} \\
{$^b$Each set of observation consists of 2 [loops]$\times$4 [half-wave plate angles]$\times$[time of integration].} \\
{$^c$The ISP-corrected Stokes parameters in the 6400--7000\,\AA\ band.} \\
\end{normalsize}
\end{center}
\end{table*}

\subsection{Interstellar Polarization}~\label{sec:specpol_isp}
The ISP along the line of sight toward SN\,2025kzr was estimated from the VLT spectropolarimetry obtained at epoch 3. At 34.0 days after explosion, the photospheric continuum polarization is expected to have declined to a negligible level as the ejecta become increasingly optically thin to electron scattering~\citep{2008ARA&A..46..433W}. This is corrobrated by the emergence of numerous iron-group element (IGE) emission lines visible in the Epoch-3 spectrum (\suppmattwo Figure~\ref{fig:iqu_ep3}). In the prototypical Type Ibn SN\,2006jc, for example, Fe\,{\sc ii} features were observed to develop in the 4500--5500\,\AA\ region $\sim 10$ days after peak brightness~\citep{2009MNRAS.392..894A}, and the He\,{\sc i}, Fe\,{\sc ii}, and Ca\,{\sc ii} emission lines strengthened with time relative to the continuum, with the apparent blue pseudocontinuum becoming increasingly dominated by numerous blended IGE lines~\citep{2007ApJ...657L.105F}. The appearance of such IGE features is a well-established spectroscopic indicator that the photosphere has receded deep into the ejecta and that the outer layers are becoming optically thin to continuum processes, rendering the Epoch-3 spectropolarimetry of SN\,2025kzr an appropriate basis for ISP estimation~\citep{2001ApJ...550.1030W, 2001ApJ...556..302H, 2012A&A...545A...7P}. 

Under this assumption, the ISP was estimated as the error-weighted mean of the Stokes $Q$ and $U$ parameters across the wavelength range 4000--8600~\AA. 
We excluded the strong telluric absorption bands (6820--6880~\AA\ and 7550--7750~\AA), and the broad Ca\,{\sc ii} near-infrared triplet (hereafter Ca\,{\sc ii}\,NIR3) spanning 7900--8500~\AA, which may carry residual intrinsic line polarization at this epoch, before computing the weighted mean~\citep{2012A&A...545A...7P}. 
The resulting ISP estimates are $Q_{\rm ISP} = 0.154 \pm 0.145$\% and $U_{\rm ISP} = 0.396 \pm 0.173$\%, which have been subtracted from all three epochs prior to our analysis of the intrinsic time-varying spectropolarimetry of SN\,2025kzr (see Figure~\ref{fig:1} ISP1). The uncertainties in these ISP estimates are not included in the quoted uncertainties of the polarization degree or position angle.

As an independent check, the Galactic ISP estimated by~\citet{deWet2026} from a bright field star in the Nordic Optical Telescope (NOT) ALFOSC field of view yields $(Q,U)_{\rm ISP} = (0.00 \pm 0.04,\ 0.33 \pm 0.04)\%$ ($P_{\rm ISP} \approx 0.33\%$, $\theta_{\rm ISP} \approx 45^{\circ}$, see Figure~\ref{fig:1} ISP2), which is consistent with our late-time estimate ($P_{\rm ISP} \approx 0.43\%$, $\theta_{\rm ISP} \approx 34^{\circ}$) within $1\sigma$. 
The field star probes only part of the Galactic dust column, whereas our estimate includes the full Galactic and host-galaxy contributions. The agreement therefore indicates the host contribution is small. This is also consistent with the low host reddening of $E(B-V)_{\rm host} \approx 0.05$ mag as estimated from the Na\,{\sc i}\,D absorption~\citep{deWet2026}.

\section{Polarization Properties of SN\,2025kzr}~\label{sec:analysis}
\subsection{Continuum Polarization}~\label{subsec:Continuum Polarization}

We corrected the normalized Stokes spectra for wavelength-dependent instrumental polarization and derived and debiased the polarization degree and PA as described by \citet{2007ASPC..364..503F}, \citet{2014A&A...561A..82S}, and \citet{2017MNRAS.464.4146C}.
We estimate the continuum polarization of SN\,2025kzr at 8, 25, and 34 days from our VLT spectropolarimetric observations, using the ISP-corrected Stokes parameters in the 6400--7000\,\AA\ band, a wavelength range expected to be largely free of strong polarized spectral features (e.g.,~\citealp{2009A&A...508..229P,Yang2020}).
The error-weighted mean Stokes parameters are reported in Table~\ref{Table_pol}. As shown in Figure~\ref{fig:1}, the measurements in this band agree with those across the full 4000--8600\,\AA\ range. The measurements in the 5100--5700\,\AA\ band, shown for comparison, are also consistent with both wavelength ranges within the uncertainties.
The intrinsic continuum polarization in the 6400--7000\,\AA\ band is \(P = 0.462\% \pm 0.071\%\) at 8\,d. It declines rapidly over the following two weeks, reaching \(P = 0.042\% \pm 0.071\%\) at \(+12\)\,d, and is consistent with zero by $+$21\,d  with \(P = 0.085\% \pm 0.182\%\).

\subsection{Temporal Evolution of Continuum Polarization}~\label{sec:specpol_cont}

The spectroscopic evolution of SN\,2025kzr proceeds through the four characteristic phases as discussed by~\citet{deWet2026}, which also provide the framework for interpreting the time-variable polarization. 
During the first $\sim10$\,d, the spectrum is dominated by emission lines arising from the interaction between the ejecta and the CSM. These features subsequently disappear as intermediate-velocity He\,{\sc i} P~Cygni profiles emerge, with absorption minima at velocities of $\sim1500\,{\rm km~s^{-1}}$. Broader ejecta features begin to appear at $\sim15.8$\,d and become prominent by $\sim23.8$\,d, marking the growing contribution of the underlying SN ejecta. By $\sim$34\,d, numerous IGE lines are visible in the spectrum.
To place the polarization evolution in this context, we combine the measurements at 4.9 and 10.9\,d reported by~\citet{deWet2026} with our new VLT measurements at 8.0, 25.0, and 34.0\,d. 
These five epochs span the transition from interaction- to ejecta-dominated emission and reveal clear evolution in the Stokes $Q-U$ plane (Figure~\ref{fig:1}).

%To facilitate a direct comparison with the $V$-band imaging polarimetry of \citet{deWet2026}, we selected the continuum polarization in a region of the $V$ band.
To compare our measurements directly with the $V$-band imaging polarimetry obtained with ALFOSC mounted on the Nordic Optical Telescope \citep[NOT;][]{deWet2026}, we integrated our continuum Stokes spectra over the $V$-band transmission function. 
As shown in Figure~\ref{fig:1}D, the continuum polarization at 4.9\,d lies close to the origin, while the spectrum is still dominated by emission lines. By 8.0\,d, the Stokes vector has moved substantially toward positive $Q$ and negative $U$, reaching $(Q,U)\approx(0.21,-0.34)\%$, coincident with the fading of the flash-ionized lines and the emergence of He\,{\sc i} P~Cygni absorption. 
At 10.9\,d, when the flash features have disappeared and the spectrum is dominated by He\,{\sc i} P~Cygni profiles, the polarization shifts to $(Q,U)\approx(0.49,0.32)\%$, manifesting a pronounced change in both polarization degree and PA. 

At 25.0 and 34.0\,d, when broader ejecta features are visible in the spectrum, the observed continuum polarization lies close to the ISP region in the $Q-U$ plane. The relatively large ISP uncertainty ($\sim$0.15\%) in each Stokes parameter prevents a precise determination of the intrinsic continuum polarization. Nevertheless, the measurements place useful upper limits on it, and hence on the global asphericity of the electron-scattering photosphere and the underlying ejecta once the emission has become ejecta-dominated.
We note that subtracting the ISP merely shifts the $Q-U$ track and does not affect the displacements between epochs. The temporal evolution of the continuum polarization therefore remains robust despite the significant uncertainty in the ISP.

\subsection{Line Polarization}~\label{sec:specpol_line}

\begin{figure*}
    \centering
    \includegraphics[width=0.95\textwidth]{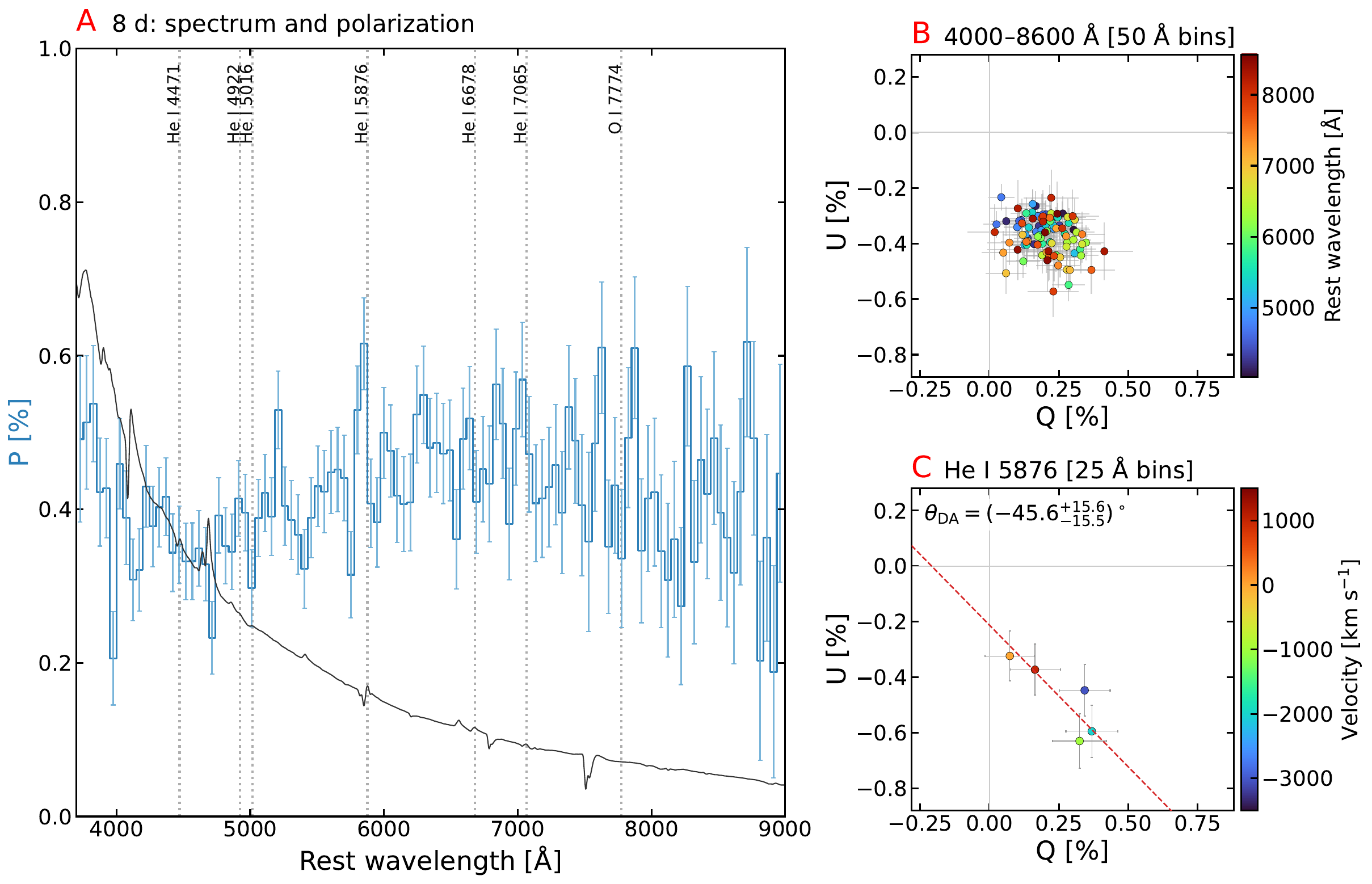}
    \caption{
    Spectropolarimetry of SN\,2025kzr at 8\,d. All measurements in Panels A--C have been corrected with the ISP as detailed in Section~\ref{sec:specpol_isp}.
    {\it Panel A:} Polarization (blue histogram) and scaled total-flux spectrum  (black curve). Major spectral features are labeled with velocity of $0$\,km\,s$^{-1}$.
    {\it Panel B:} Polarization across 4000--8600\,\AA, with the wavelength of each 50\,\AA\ bin indicated by the color bar. 
    {\it Panel C:} Velocity-resolved Stokes parameters across He\,{\sc i}\,$\lambda5876$, measured in 25~\AA\ bins and color-coded by velocity relative to the line rest wavelength. 
    The red dashed line shows the linear fit to the binned measurements. The slope and uncertainties are indicated by the text.
    }~\label{fig:2}
\end{figure*}

At $+8$\,d, SN\,2025kzr exhibits continuum polarization across the observed wavelength range (Figure~\ref{fig:2}A). In the $ Q-U$ plane, the continuum measurements form a compact cluster offset from the origin (Figure~\ref{fig:2}B), indicating a net continuum polarization with little variation in PA with wavelength. In contrast, the velocity-resolved measurements across He\,{\sc i}\,$\lambda5876$ define an approximately linear sequence (Figure~\ref{fig:2}C). We fit these line measurements with a dominant axis of the form
\begin{equation}
     U =\alpha + \beta Q, 
\end{equation}
where $\alpha$ and $\beta$ are respectively the intercept and slope of the error-weighted linear fit.
We obtained $R^2=0.695$ and $\chi^2/\mathrm{DoF}=0.411$ over $-3500 \leq v \leq 1500\ \mathrm{km\,s^{-1}}$, indicating a possible but statistically inconclusive alignment.
The resulting dominant-axis angle is
$\theta_{\rm DA}=(-45.6^{+15.6}_{-15.5})^\circ$. The PA derived from the continuum Stokes vector is similar to that across the He\,{\sc i}\,$\lambda5876$ P~Cygni profile, suggesting a common preferred orientation, in projection, for the continuum-scattering region and the He\,{\sc i}\,$\lambda5876$ opacity.
Departures from a single dominant axis would instead indicate contributions from multiple geometrical components or departures from axisymmetry~\citep{2003ApJ...591.1110W, 2010ApJ...722.1162M}. Within the uncertainties, the He\,{\sc i}\,$\lambda5876$ polarization at 8\,d is consistent with a single dominant axis, and hence with an axisymmetric distribution of the line-forming material.

The total-flux spectra also show P~Cygni profiles in He\,{\sc i}\,$\lambda6678$ and $\lambda7065$, similar to that of $\lambda5876$; however, none of the three He\,{\sc i} features, including $\lambda5876$, shows a clearly resolved line-polarization profile above the noise. 
The dominant axis inferred from the velocity-binned $\lambda5876$ measurements is therefore tentative.
Its agreement with the continuum PA (see, e.g., the orientation of the data cloud in Figure~\ref{fig:2}B) hints at a shared projected orientation, but the He\,{\sc i} polarization data alone do not provide a firm geometric constraint.

\section{Temporal Evolution Model of Continuum Polarization}~\label{sec:model}

As displayed in Figure~\ref{fig:1}, the temporal evolution of the continuum polarization traces a rotating trajectory in the $Q-U$ plane. Such a rotation arises naturally from a misaligned ejecta-CSM geometry, whereas a coaxial axisymmetric configuration would generally evolve along a single common axis in the $Q-U$ plane (see, e.g.,~\citealp{Yang2025}). 
Our measurements span the period during which the shock sweeps through the confined, dense CSM and the emission transitions to being ejecta-dominated, and therefore provide a sensitive probe of the changing geometry.  
We use ellipsoidal models to constrain the geometry in each phase, and hydrodynamic simulations to illustrate how the proposed dynamics may give rise to the observed rotation of the continuum polarization in the $Q-U$ plane. 
Because these models assume simplified, global ellipsoidal configurations for both the ejecta and the CSM, and because of the degeneracies among the model parameters, they are not intended to uniquely reconstruct the 3D structure, but rather to characterize its plausible geometric evolution.

\subsection{Monte Carlo Constraints on CSM and Cold Dense Shell Geometry}~\label{sec:3.1}

\begin{figure*}
    \centering
    \includegraphics[width=\textwidth]{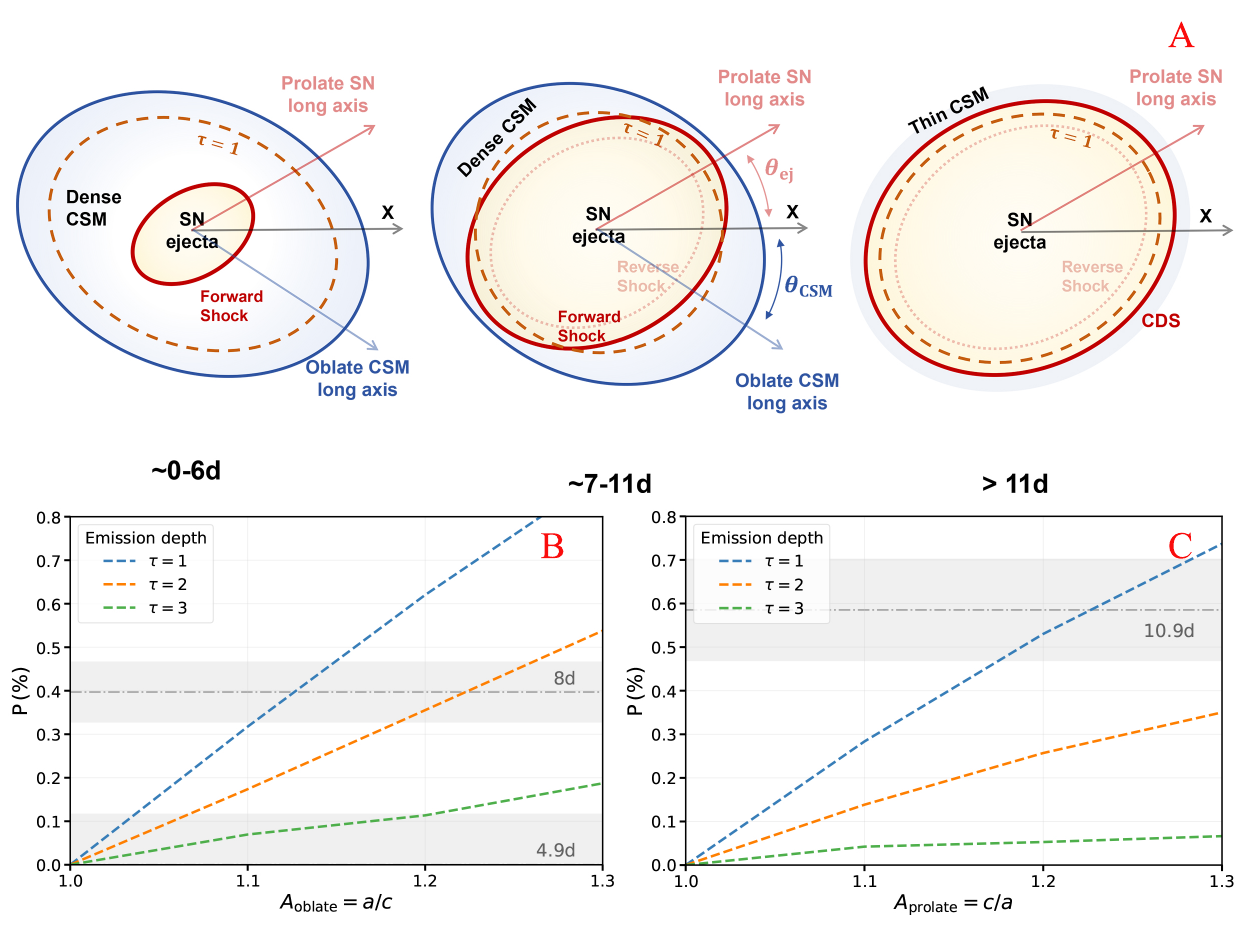}
\caption{Time-variable ejecta-CSM geometry and the model continuum polarization. 
\textbf{Panel A:} Schematic of the interaction region that spans the transition from CSM-interaction-dominated to CDS-dominated phases (see Section~\ref{sec:specpol_cont}) as seen perpendicular to the $x$--$z$ plane.
At $\sim$0--6\,d, the forward shock (solid red) is embedded in the optically thick, oblate CSM (blue) and lies inside the $\tau=1$ surface (dashed orange).
At $\sim$7--11\,d, the forward shock approaches the outer, optically thinner part of the dense CSM.
After $\sim$11\,d, the shock has swept through the dense CSM, the CDS has formed between the forward and reverse shocks, and the $\tau=1$ surface lies within the CDS at some angles.
The long axes of the prolate SN ejecta (red arrow) and the oblate CSM (blue arrow) are misaligned, with inclinations $\theta_{\rm ej}$ and $\theta_{\rm CSM}$
(Section~\ref{sec:model}). The schematic is not to scale. 
\textbf{Panel B:} Model continuum polarization $P$ of the oblate, CSM-dominated phase as a function of the axis ratio $A_{\rm oblate}=a/c$, where $a$ and $c$ are the equatorial and polar semiaxes. 
Blue, orange, and green dashed curves show photon packets injected at electron-scattering optical depths of $\tau=1$, 2, and 3, respectively.
%The gray dash-dotted line and 
Gray-shaded band shows the observed, ISP-corrected and bias-corrected polarization and its $1\sigma$ uncertainty at 4.9\,d and 8d, respectively. 
\textbf{Panel C:} Same as Panel~B but for the prolate, CDS-dominated phase as a function of $A_{\rm prolate}=c/a$, compared with the observation at 10.9\,d.
Panels B and C show that polarization increases with asphericity and decreases with injection depth.
}~\label{fig:3}
\end{figure*}

We use the 3D Monte Carlo polarization code MCPol~\citep{Wen2023} to model the continuum polarization, following methods widely used in SN polarization studies \citep{Hoflich1991, Kasen2003, 2020ApJ...902...46Y, 2023MNRAS.520..560H}. 
We adopt an oblate ellipsoid for the immediate CSM, which produces the early spectral signatures of interaction, and a prolate ellipsoid for the cold dense shell (CDS) formed as the shock propagates through the CSM.
The continuum polarization depends on both the geometry and the optical depth of the scattering region. Consequently, a low polarization degree can result either from a nearly spherical configuration or from depolarization by multiple scattering at high optical depth (see, e.g.,~\citealp{Hoflich1991, 2020ApJ...902...46Y, 2023MNRAS.520..560H}). 
Around the bolometric light curve peak (4--6\,d), the spectrum indicates ongoing ejecta-CSM interaction, with optically thick CSM lying above the forward shock. 
This is supported by the persistent flash-ionized lines with broad electron-scattering wings and by the peak of the bolometric light curve at $\sim$5.8\,d, which has been interpreted as shock breakout from dense CSM~\citep{deWet2026}. Multiple scattering in such a medium efficiently depolarizes the emergent continuum. 
We therefore interpret the low polarization measured at 4.9\,d primarily as an optical-depth effect rather than as evidence for a nearly spherical CSM (Figure~\ref{fig:3}A). Within our MCPol model, this interpretation is consistent with a moderately aspherical CSM with an axis ratio of $\sim$1.2 (Figure~\ref{fig:3}B).

As the shock approaches the outer dense CSM (around 8\,d), the optical depth above the emitting surface decreases, so the aspherical CSM can produce a more pronounced polarization signal. 
Because the observed polarization is dominated by the last few scattering events, we assume an optical depth of $\tau\approx 1$--2 above the photon-emitting surface. 
Under this assumption, the continuum polarization constrains the CSM to an oblate geometry with an axis ratio of \(\sim1.1\)–1.25 (Figure~\ref{fig:3}B).
Independently, the He\,{\sc i}\,$\lambda$5876 line polarization defines a dominant axis corresponding to a sky PA of $\theta_{\rm CSM} = \theta_{\rm DA}\approx23^{\circ}$, which traces the preferred orientation of the He-rich, unshocked CSM projected on the plane of the sky (see Figure~\ref{fig:3}A).

%By $\sim$11\,d, the high-ionization flash features have disappeared~\citep{deWet2026}, indicating that the forward shock has swept through the dense inner CSM, which for a typical shock velocity of $\sim 10^{4}$\,km\,s$^{-1}$ corresponds to a radius of $\sim10^{15}$\,cm.
By \(\sim11\) d, the high-ionization flash features have disappeared \citep{deWet2026}, indicating that the forward shock has swept through the dense inner CSM, which corresponds to a radius of $\sim10^{15}$\,cm for a typical shock velocity of $\sim 10^{4}$\,km\,s$^{-1}$. 
The shock may also have reshaped the global geometry, producing a prolate CDS. As the photosphere recedes into the CDS, the observed continuum emission emerges from within the shell, so the continuum polarization is dominated by the CDS geometry. 
Because the CDS density increases steeply inward, the optical depth is larger in its inner regions.
Consequently, the continuum polarization is primarily sensitive to the geometry of the outer CDS layers near the photosphere (\(\tau\sim1\)).
Under this assumption, our model favors a prolate CDS with an axis ratio of \(\sim1.2\)–1.25 (Figure~\ref{fig:3}C).

As the CDS expands and cools, it becomes increasingly optically thin, allowing the photosphere to recede into the ejecta. Continued expansion and recombination further reduce the electron-scattering optical depth of the ejecta, which weakens the polarization produced by scattering. Consequently, the continuum polarization declines and falls below our detection threshold.

\subsection{Hydrodynamic Model and $Q-U$ Evolution}

To investigate the temporal evolution of the continuum polarization, we
constructed a radiation-hydrodynamic model of the ejecta-CSM interaction.
We followed the hydrodynamic and radiative evolution with the 3D radiation-hydrodynamic code \texttt{Athena++}~\citep{2020ApJS..249....4S, 2021ApJS..253...49J}
and post-processed the resulting structures with MCPol to calculate the time-dependent Stokes parameters (see Appendix~\ref{sec:Athena}).  
This simulation is intended to illustrate how the observed evolution of the continuum polarization can plausibly arise from the ejecta–CSM interaction. It is not designed to provide a unique or quantitative fit to the data, nor to constrain the physical parameters of the system. 

Following the constraints derived in Section.~\ref{sec:3.1}, we model the SN ejecta as a prolate ellipsoid with an axis ratio of 1.25, expanding from within the CSM along an axis oriented at \(\theta_{\rm ej}=+8^\circ\). This orientation is consistent with the continuum PA measured at $+10.9$\,d.
The CSM is represented by an oblate ellipsoidal density distribution with an axis ratio of 1.2 and a symmetry axis at \(\theta_{\rm CSM}=-25^\circ\). 
Both angles are measured in the \(x\)–\(z\) plane from the \(+x\)-axis, as viewed from \(+y\), with counterclockwise rotations defined as positive.
%Both angles are measured in the \(x\)–\(z\) plane, clockwise from the \(z\)-axis as viewed from the \(+y\) direction. 
These angles are guided by the observed polarization position angles and represent projected orientations on the plane of the sky, rather than direct observables of the three-dimensional symmetry axes. This misalignment produces an asymmetric interaction region and an anisotropic radiation field, which together provide the conditions required for time-variable continuum polarization.

\begin{figure*}
    \centering
    \includegraphics[width=\textwidth]{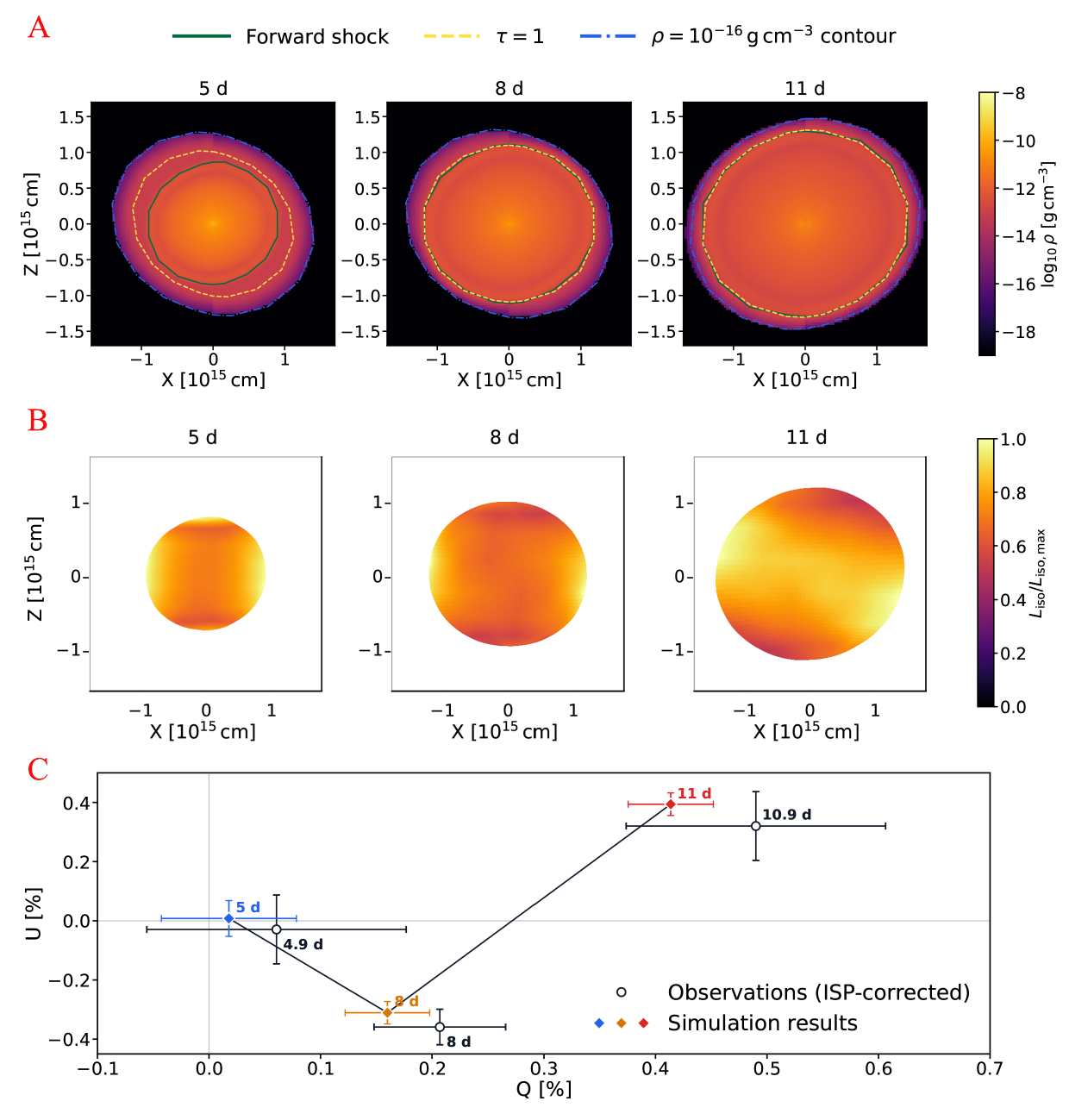}
    \caption{Hydrodynamic structure, emitting-surface brightness, and polarization evolution of the model at 5, 8, and 11\,d. 
    {\it Panel A:} Density maps in the $x$--$z$ plane. Colors show $\log_{10}(\rho/\mathrm{g\,cm^{-3}})$. The dark-green solid, yellow dashed, and blue dash-dotted curves denote the forward shock, the $\tau=1$ surface, and the \(\rho=10^{-16}\,\mathrm{g\,cm^{-3}}\) contour tracing the CSM geometry, respectively.
    {\it Panel B:} Normalized brightness distributions of the emitting surfaces viewed from the $-y$ direction. The 5\,d surface is the shock surface, whereas the 8 and 11\,d surfaces are defined by $\tau=1$. Colors indicate $L_{\rm iso}/L_{\rm iso,max}$, normalized independently at each epoch. 
    {\it Panel C:} Comparison between the modeled and ISP-corrected observed Stokes parameters in the $Q-U$ plane. Colored diamonds show the $+y$ model predictions at 5, 8, and 11\,d; open circles present the observations at 4.9, 8.0, and 10.9\,d. The black solid curve connects the model epochs, and dotted segments illustrate the model--observation offsets. Error bars indicate $1\sigma$ uncertainties, including Monte Carlo uncertainties for the model and measurement uncertainties for the observations.}
    \label{fig:4}
\end{figure*}

Figure~\ref{fig:4} shows the hydrodynamic structure and the polarization evolution of the model at 5, 8, and 11\,d. Panel~A presents density slices in the \(x\)–\(z\) plane, illustrating the forward shock as it propagates through and emerges from the dense, aspherical CSM, followed by the formation of the CDS. 
The \(\tau=1\) surface indicates the transition between optically thick and optically thin regions, while the \(\rho=10^{-16}\,\mathrm{g\,cm^{-3}}\) isodensity contour traces the outer boundary of the dense CSM. 
At 5\,d, the forward shock remains interior to the \(\tau=1\) surface, so we take the shock surface to be the effective photon-packet emission surface at this epoch. 
By 8\,d, the shock has propagated to the outer, optically thin layers of the CSM. As the shock continues to expand and the optical depth decreases, the \(\tau\approx1\) surface recedes into the CDS and lies within it by 11\,d. 
We therefore adopt the \(\tau\approx1\) surface as the effective photon-packet emission surface  at 8 and 11\,d, during the CDS-dominated phase. 

Figure~\ref{fig:4}B shows the normalized luminosity distribution over the photon-packet emission surface. The anisotropic density distribution produces enhanced emission from the denser regions, so the surface luminosity is nonuniform and evolves with the geometry of the interaction region.
This nonuniformity in turn yields an anisotropic distribution of emitted photon packets. 

For each epoch, we propagated \(2\times10^{6}\) photon packets and extracted the emergent Stokes parameters for the \(+y\) viewing direction  (see Figure~\ref{fig:4}C). The model evolves from nearly zero polarization at 5\,d to negative \(U\) at 8\,d, and then to positive \(Q\) and \(U\) at 11\,d,  qualitatively consistent with the observed polarization evolution. 
This rotation in the $Q-U$ plane reflects a transition from polarization dominated by the CSM geometry to polarization dominated by the CDS geometry, with the nonuniform luminosity distribution over the emission surface also contributing.

\section{Discussion}~\label{sec:discussion}

The continuum polarization of SN\,2025kzr traces the evolving geometry of the ejecta-CSM interaction. Around the bolometric light-curve peak ($\sim5$\,d), multiple scattering in the optically thick CSM can suppress the net polarization. As the shock approaches the outer boundary of the CSM ($\sim8$\,d), the $Q-U$ vector becomes increasingly sensitive to the asphericity and projected orientation of the CSM.
Once the dense CSM has been swept up, the CDS increasingly shapes the continuum polarization. At later times, as the photosphere recedes into the ejecta and the electron-scattering optical depth declines, the polarization weakens. 
The observed evolution thus reflects the changing contributions of the CSM and CDS, followed by a declining polarization signal from the underlying ejecta.

\subsection{The Geometry and Origin of the CSM}

The polarization reveals a preferred projected axis and indicates that the CSM is likely nonspherical. Such a configuration can arise in binary-interaction scenarios. In nonconservative Roche-lobe overflow, a large fraction of the transferred mass can be lost through the outer Lagrange point (L2) when the mass-transfer rate is high, forming an equatorially concentrated circumbinary outflow~\citep{2023MNRAS.519.1409L}. 
In common-envelope evolution, 3D simulations show that the rapid inspiral of the companion can transform the spherical envelope into a disk/torus-like circumbinary structure~\citep{2024A&A...691A.244V}.
For the Type Ibn SN\,2023fyq, mass transfer from a helium star onto a compact companion has been invoked to form an equatorial disk of $\sim$0.6\,$M_{\odot}$, the interaction of which with the SN ejecta powers the second light-curve peak~\citep{2024ApJ...977..254D}. More generally, binary-evolution models show that mass transfer before core collapse can form a circumbinary disk with masses and radii compatible with those inferred for many Type Ibn SNe~\citep{2025A&A...696A.103E}. Such a disk-like CSM configuration can qualitatively reproduce their light-curve morphology. The geometry of this material at core collapse, however, remains unconstrained.
These binary scenarios therefore offer plausible ways to account for the inferred CSM distribution; however, the polarization geometry alone cannot identify a unique formation channel. A single massive progenitor, as favored by~\citet{deWet2026} on the basis of the precursor timescale and energetics, therefore cannot be excluded.

An alternative possibility is that the precursor arose from wave-driven mass loss in a massive Wolf–Rayet progenitor, with waves excited by late-stage nuclear burning depositing energy in its outer layers and driving an outburst \citep{deWet2026,2012MNRAS.423L..92Q,2014ApJ...780...96S, Fuller2017}. 
The precursor of SN\,2025kzr, detected $\sim$55\,d before explosion, is consistent with such an energetic mass-loss episode \citep{deWet2026}; however, the angular distribution of the resulting outflow remains poorly constrained.
Eruptive mass loss need not produce a spherical CSM. The prototypical Type Ibn SN\,2006jc had a luminous precursor outburst about two years before core collapse~\citep{2007Natur.447..829P, 2007ApJ...657L.105F}. Its double-peaked He\,{\sc i} emission profiles were interpreted as evidence for an aspherical He-rich CSM \citep{2007ApJ...657L.105F}. This provides an observational precedent for a nonspherical CSM associated with an eruption; however, it does not identify the mechanism that produced the asymmetry. 

Polarization facilitates the discrimination between line-forming geometries. In the $Q-U$ plane, an approximately axisymmetric structure is expected to produce data points aligned along a dominant axis. A 3D clumpy distribution can instead produce velocity-dependent rotations of the PA and loop-like trajectories~\citep{Tanaka2017}. 
If the He\,{\sc i}\,$\lambda$5876 polarization of SN\,2025kzr is aligned along a dominant axis with no significant loop, this would favor a coherent, large-scale projected asymmetry, as randomly oriented clumps would manifest small-scale asymmetries and thus not produce such a pattern  indicative of axisymmetry.
We note that at the present signal-to-noise ratio (S/N), the He\,{\sc i} measurements are only tentative. They do not exclude smaller-scale structure superposed on global asymmetry.

\subsection{Evidence for Misaligned Ejecta and CSM Symmetry Axes}

The prolate ejecta geometry adopted in our illustrative model raises the possibility that the explosion itself was axisymmetric and inherently misaligned with the CSM. 
Large-scale asymmetries can develop through hydrodynamic instabilities during core collapse, as nonradial perturbations of the stalled accretion shock would grow into turbulence~\citep{2003ApJ...584..971B}, and multidimensional convection and instabilities shape the explosion engine (see, e.g., \citealp{2020LRCA....6....3M}).
While rapid rotation and magnetic fields may produce more strongly axial outflows, as in simulations where magnetic stresses launch a magnetohydrodynamic (MHD) jet~\citep{1999ApJ...524L.107K, 2007ApJ...664..416B}. 
Polarimetric observations have revealed related behavior in other events. For example, spectropolarimetry of SN\,2005bf was interpreted in terms of a jet-like structure tilted with respect to the symmetry axis of the photosphere~\citep{2007MNRAS.381..201M, 2009ApJ...699.1119T}. In the Type II SN\,2024ggi, the later H-rich ejecta share the breakout axis, while the CSM-interaction axis differs~\citep{Yang2025}. These events show that polarization can preserve information about distinct projected axisymmetric configurations within an exploding massive star.
We note that these events span different SN subtypes (Type Ib/c, Type II, and Type Ibn). It remains unconstrained whether they share a common explosion mechanism.

In our model, the misalignment between the prolate ejecta and oblate CSM produces an evolving interaction region. As the shock propagates through the CSM, the CDS becomes an increasingly important source of continuum polarization. This provides a plausible explanation for the observed $Q-U$ evolution. The rotation of the PA, however, requires only that the polarized emitting components be noncoaxial, not that the explosion itself be intrinsically aspherical. For example, an aspherical CSM alone can generate an asymmetric interaction region around initially spherical ejecta~\citep{2019ApJ...887..249S}, and a misalignment could then arise between components of the CSM or of the interaction zone.
Hydrodynamic calculations show that aspherical CSM can also generate an asymmetric interaction region around initially spherical ejecta \citep{2019ApJ...887..249S}. 

Our simulation demonstrates the plausibility of the proposed geometric evolution during the interaction-dominated phase but does not uniquely determine the ejecta shape or explosion mechanism. It assumes smooth, axisymmetric ellipsoids and calculates electron-scattering polarization in post-processing from prescribed emission surfaces, without self-consistently modeling radiation transport associated with a specific explosion mechanism. Degeneracies among geometry, viewing angle, and optical depth, together with uncertainty in the ISP, preclude a unique three-dimensional reconstruction.
Future progress will require both better observations and better models. Denser spectropolarimetric time series, extending to earlier epochs, when the CSM is still optically thick, as well as to the nebular phases, would constrain how the geometry changes with depth. Higher S/N would allow the He\,{\sc i} line polarization to be resolved, and an independent ISP estimate would anchor the absolute polarization levels and PAs. On the model effort, self-consistent 3D radiation-hydrodynamic models that integrate polarized radiative transfer to the dynamics, include line and frequency-dependent opacities, and explore the tilt angle, inclination, and clumpiness would test whether the proposed geometry is unique. 
%Our simulation thus demonstrates only the plausibility of the proposed geometric evolution during the interaction-dominated phase, and it does not uniquely determine the ejecta shape or  the explosion mechanism. Its simplified treatment of radiation, together with degeneracies among geometry, viewing angle, optical depth, and opacity, 
%{\yy Perhaps slightly expand the text that says the caveats: There are several simplifications that limit the model of SN\,2025kzr. (i) The ejecta and CSM are idealized as smooth, axisymmetric ellipsoids, with no clumping or density substructure; (ii) The polarization is computed in post-processing. In particular, photon packets are emitted from a prescribed surface, which is the shock surface at 5\,d and the $\tau\simeq1$ surface at later epochs. An optical depth above this surface is assumed, instead of being obtained from a self-consistent radiative transfer calculation; (iii) The calculation only accounts for electron scattering, so the model provides no constraint on the line polarization or wavelength dependence; (iv) The geometry, viewing angle, optical depth, and opacity are partly degenerate; (v) The ISP uncertainty limits the determination of the polarization level intrinsic to the SN. These factors'} precludes a unique 3D reconstruction. 

Finally, we note that evidence for noncoaxial ejecta and CSM is emerging in other CCSNe. In the Type II SN\,2023ixf, the polarization PA rotates as the ejecta emerge from asymmetric CSM~\citep{2023ApJ...955L..37V, 2026ApJ..1000...18V}, and in SN\,2024ggi the shock breakout and the ejecta-CSM interaction manifest misaligned symmetry axes~\citep{Yang2025}. The transitional Ibn/Icn SN\,2024abvb also shows a rotation of the polarization PA~\citep{2026A&A...710A..35A}. Type IIn SNe present a mixed picture. In SN\,1997eg, the narrow-line polarization axis, which traces a disk-like CSM, differed by $\sim12^{\circ}$ from that of the elongated ejecta~\citep{2008ApJ...688.1186H}. Most other Type IIn SNe show a persistent PA, as in the sample of \citet{2024MNRAS.529.1104B} and in SN\,2017hcc~\citep{2024MNRAS.527.6090M}. One should remain cautious about this statement on Type IIn events since their polarization mainly traces the CSM. A persistent angle offers limited capability to test whether the ejecta share its axis. SN\,2010jl likewise has a substantially aspherical configuration, but only single-epoch spectropolarimetry~\citep{2011A&A...527L...6P}, which cannot constrain the relative orientation. For the SN\,1997cy-like (SN~Ia-CSM) event SN\,2018evt, the symmetry axes of the aspherical CSM and the SN instead appear similar~\citep{2023MNRAS.519.1618Y}. Applying time-resolved spectropolarimetry across the transition from interaction- to ejecta-dominated emission to a larger sample of CCSNe would test whether misaligned ejecta and CSM geometries are common or depend on SN type, progenitor, or CSM properties.

\begin{acknowledgments}
We thank the European Organisation for Astronomical Research in the Southern Hemisphere (ESO) for generous allocations of observing time. 
We appreciate the staff of the Paranal Observatory for their proficient and diligent support of this project in service mode. 
%Spectropolarimetry presented in this study is based on observations collected at ESO’s La Silla Paranal Observatory under program ID 0115.D-2096 [PI Y. Yang] and can be accessed via: \url{https://archive.eso.org/cms.html}. 
%{IRAF} is distributed by the National Optical Astronomy Observatories, which are operated by the Association of Universities for Research in Astronomy, Inc., under cooperative agreement with the NSF.
PyRAF, PyFITS, and STSCI$\_$PYTHON are products of the Space Telescope Science Institute (STScI), which is operated by the Association of Universities for Research in Astronomy, Inc., under NASA contract NAS5-26555. This research has made use of NASA's Astrophysics Data System Bibliographic Services, the SIMBAD database, operated at CDS, Strasbourg, France, and the NASA/IPAC Extragalactic Database (NED) which is operated by the Jet Propulsion Laboratory, California Institute of Technology, under contract with NASA. 

X.W.'s research is supported by the National Natural Science Foundation of China (NSFC grant 12603078).
Y.Y.'s research is supported by the Tsinghua University Dushi Program. 
M.\,B. acknowledges the Department of Physics and Earth Science of the University of Ferrara for financial support through grant FIRD 2025. 
%The work of A.\,Cikota is supported by NOIRLab, which is managed by the Association of Universities for Research in Astronomy (AURA) under a cooperative agreement with the NSF. 
A.V.F. is grateful for the hospitality of the Hagler Institute for Advanced Study as well as the Department of Physics and Astronomy at Texas A\&M University during part of this investigation. A.V.F. acknowledges financial assistance from many  donors. 
%S.S. is partially supported by LBNL Subcontract 7707915.
SdW is supported by a research grant (VIL60862) from VILLUM FONDEN.

\end{acknowledgments}

\section*{Data Availability}
Spectropolarimetry presented in this study is based on observations collected at ESO’s La Silla Paranal Observatory under program ID 0115.D-2096 [PI Y. Yang] and can be accessed via  \url{https://archive.eso.org/cms.html}. 
The reduced data underlying this article will be shared on reasonable request to the corresponding author. 

\begin{contribution}
X.W. and Y.Y. wrote the manuscript. 
X.W. developed the 3D Monte Carlo Polarization Simulation Code, carried out the modeling of the spectropolarimetry data, and contributed to the data interpretation. 
Y.Y. is the PI of the VLT spectropolarimetry proposal, initiated the project, carried out the observations, reduced the data, conducted spectropolarimetric analysis, and contributed to the data interpretation. 
L.W., J.C.W., and A.V.F. contributed to the proposal, data interpretation, and manuscript. 
C.A helped with the observations.
S.~D. contributed to the analysis of the NOT/ALFOSC imaging polarimetry data.  
A.V.F, J.R.M, M.B, S.~D, and S.~S provided comments on the manuscript. 
\end{contribution}

\facility{VLT:Antu (FORS2)}

\software{astropy~\citep{2013A&A...558A..33A, 2018AJ....156..123A, 2022ApJ...935..167A},  
          PyRAF~\citep{2012ascl.soft07011S}, 
L.A.Cosmic~\citep{2001PASP..113.1420V}, 
Athena++~\citep{2020ApJS..249....4S}, 
MCPol~\citep{Wen2023}, 
numpy~\citep{2020Natur.585..357H}, 
scipy~\citep{2020NaMet..17..261V}, 
matplotlib~\citep{2007CSE.....9...90H}.
          }

\setcounter{figure}{0}
\setcounter{table}{0}
\appendix
\section{Appendix}~\label{sec:Appendix}
\subsection{Spectropolarimetry of SN~2025kzr}~\label{sec:appendix_data}
\begin{figure*}[ht!]
    \centering
    \includegraphics[trim={0.0cm 0.0cm 0.0cm 0.0cm},clip,width=0.6\textwidth]{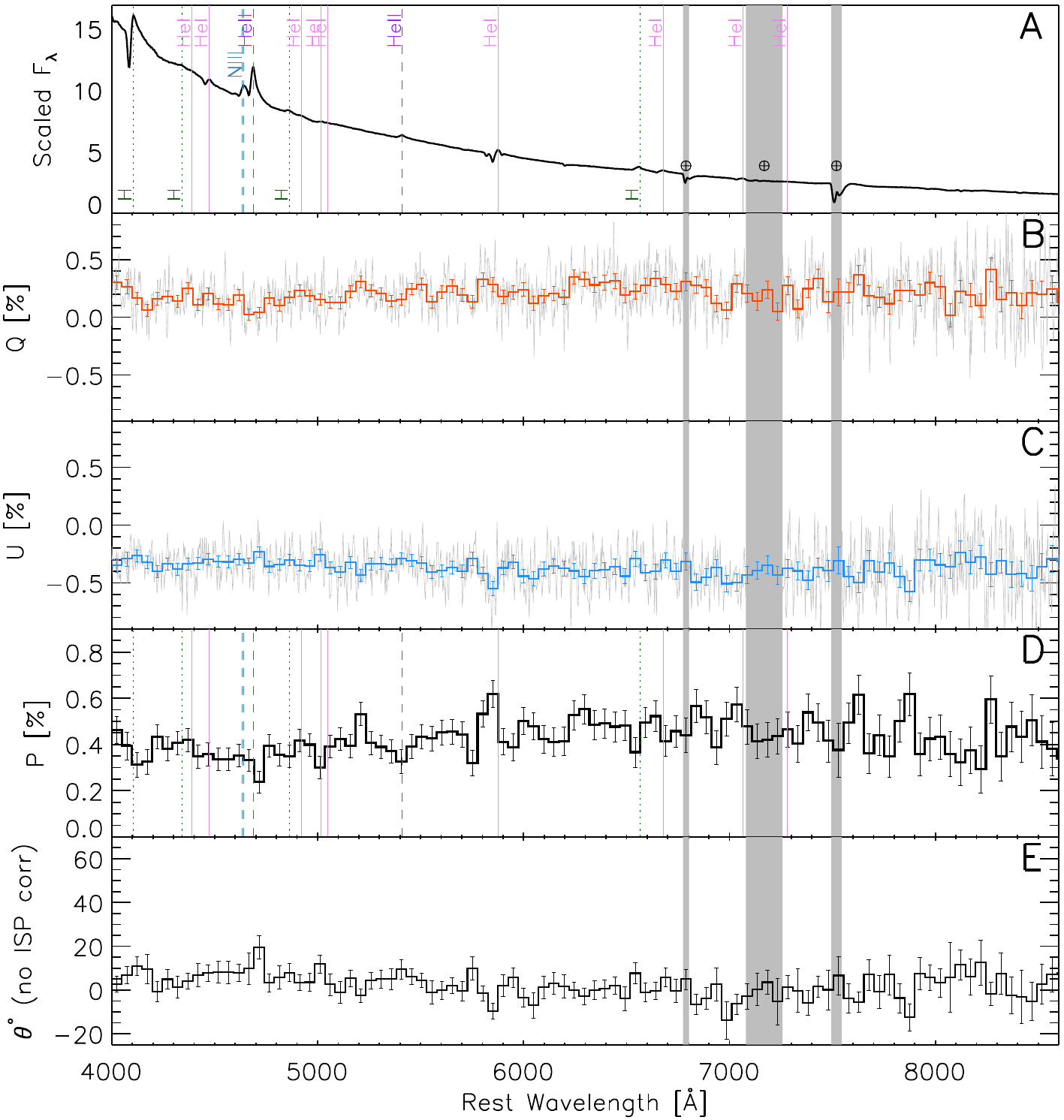}
    \vspace{-0.0 cm}
    \captionsetup{name=\suppmattwo Figure }
    \caption{Spectropolarimetry of SN\,2025kzr at 8\,d after explosion (Epoch~1). From top to bottom, the five panels show  (A) the arbitrarily scaled total-flux spectrum with major spectral features identified; (B, C) the intensity-normalized Stokes parameters $Q$ and $U$, respectively; (D) the polarization spectrum ($p$); and (E) the polarization PA. Panels (B)–(D) are corrected for the ISP, whereas Panel (E) is shown before ISP subtraction. The light-gray vertical bands identify regions of telluric contamination. All polarization spectra Panels (B)--(E) are rebinned to 50~\AA\ for clarity.}
~\label{fig:iqu_ep1}
\end{figure*}

\begin{figure}[ht!]
   \begin{minipage}[t]{0.48\textwidth}
     \centering
     \includegraphics[width=1.0\linewidth]{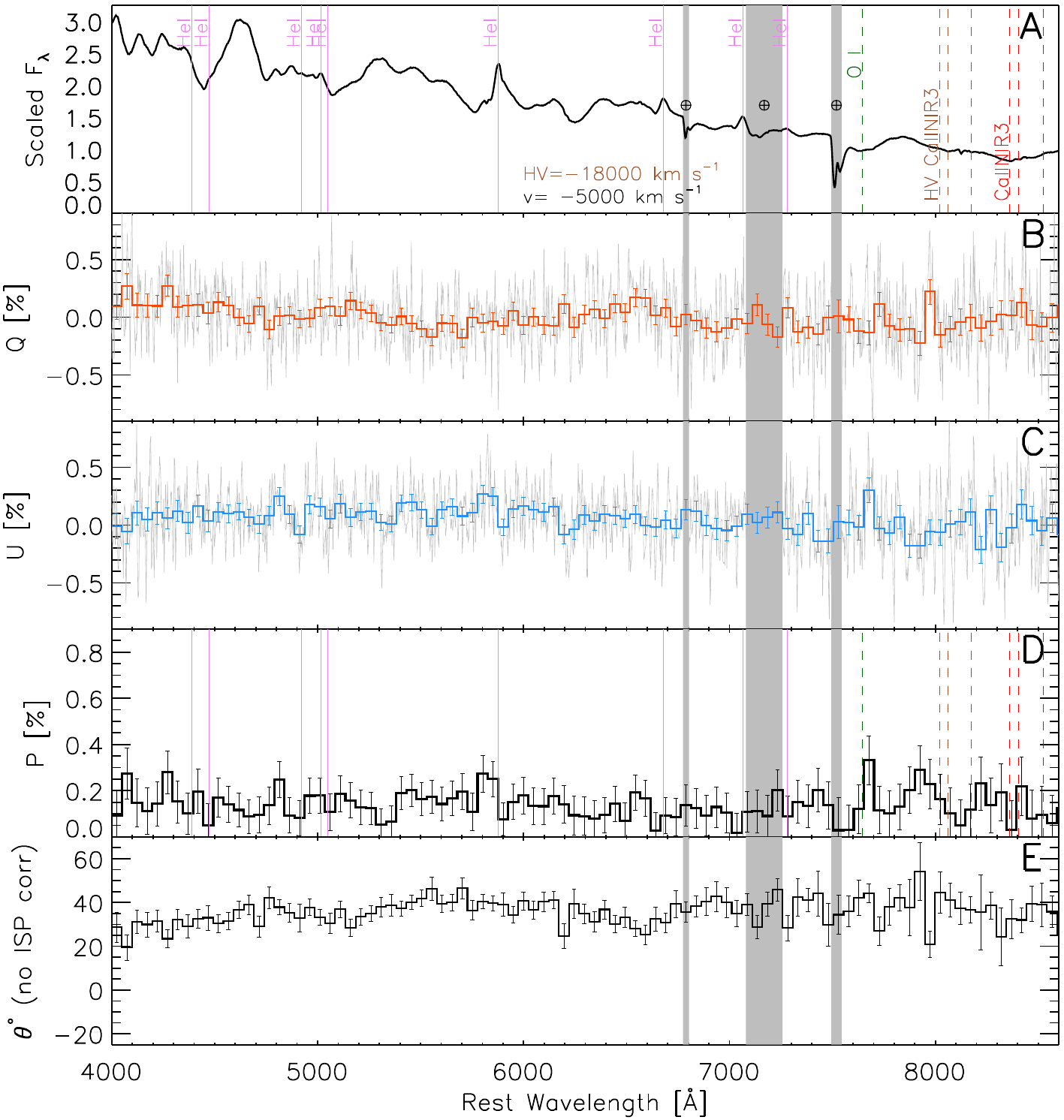}
     \captionsetup{name=\suppmattwo Figure }
     \caption{Same as \suppmattwo Figure~\ref{fig:iqu_ep1} but for 25.0\,d (Epoch~2).
     }~\label{fig:iqu_ep2}
   \end{minipage}\hfill
   \begin{minipage}[t]{0.48\textwidth}
     \centering
     \includegraphics[width=1.0\linewidth]{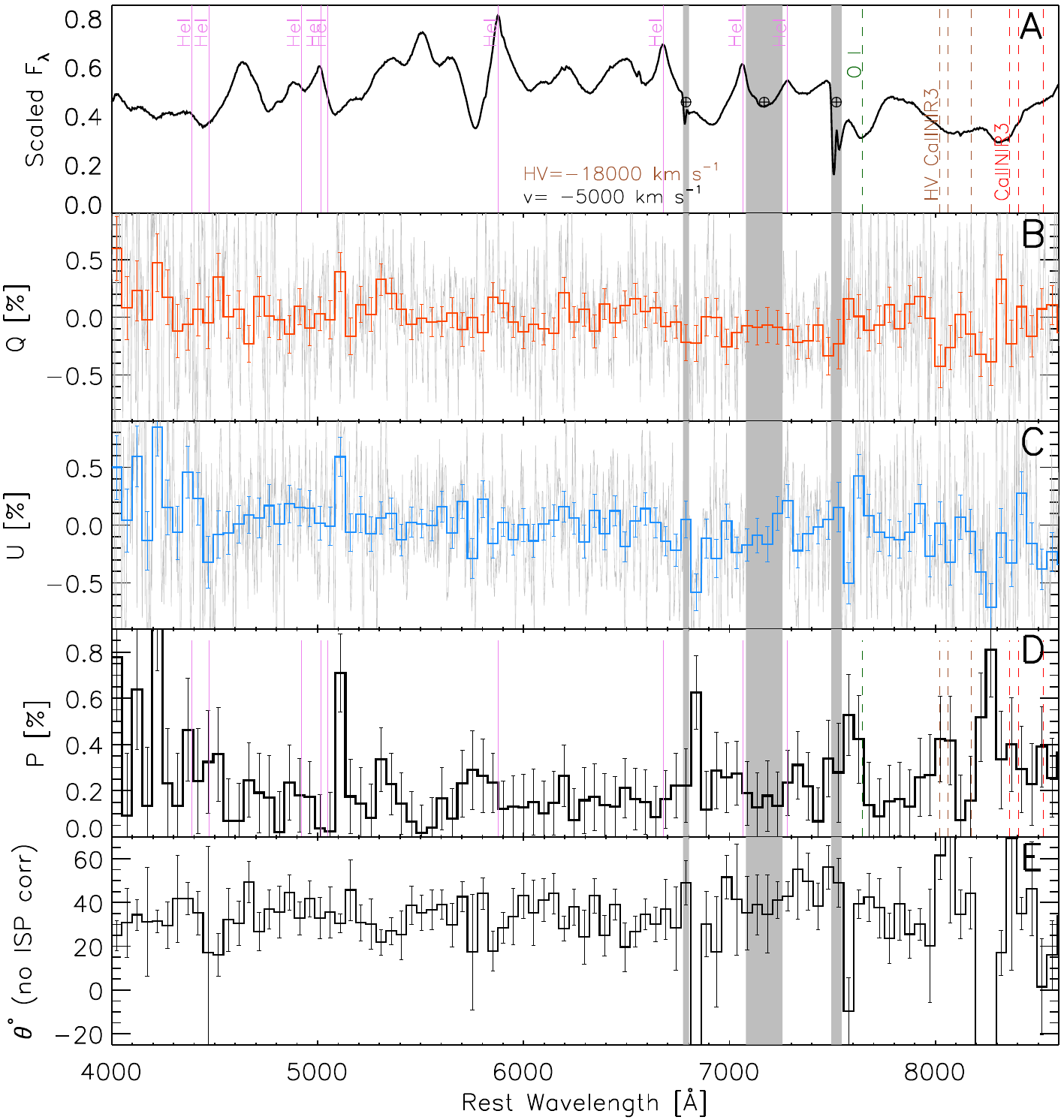}
     \captionsetup{name=\suppmattwo Figure }
     \caption{Same as \suppmattwo Figure~\ref{fig:iqu_ep1} but for 34.0\,d (Epoch~3).}~\label{fig:iqu_ep3}
   \end{minipage}
\end{figure}

We obtained spectropolarimetric observations of SN\,2025kzr at three epochs spanning the early evolution of the SN: 8.0, 25.0, and 34.0 days after the estimated time of explosion at MJD 60816.0$\pm$0.1~\citep{deWet2026}.
These data provide wavelength-resolved measurements of the total flux and linear polarization, allowing us to trace the evolution of both continuum and line polarization.
At each epoch, we extracted the total-flux spectrum, Stokes parameters $Q$ and $U$, the polarization degree $P$, and polarization PA $\theta$. 
To increase the S/N, we rebinned the polarized spectra to 50~\AA. The individual epochs are presented in \suppmattwo Figures~\ref{fig:iqu_ep1}--\ref{fig:iqu_ep3}. Regions affected by major telluric features are masked by gray color-shaded bands.
The Stokes parameters and the polarization degree are corrected for ISP, whereas the PA is shown before ISP subtraction. This is because for any source that has low continuum polarization, the polarization PA $\theta=1/2 \tan^{-1}(U/Q)$ will display random values when $Q$ is low after a baseline ISP has been subtracted.

\subsection{Three-dimensional Radiation-Hydrodynamic Polarization Simulation}~\label{sec:Athena}
We modeled the interaction between the SN ejecta and aspherical CSM with a customized 3D radiation-hydrodynamic implementation in \textsc{Athena++}~\citep{2020ApJS..249....4S,2021ApJS..253...49J}.
The calculations solve the gray radiation-hydrodynamic equations in conservative form. Our fiducial calculation employed a 3D misaligned ellipsoidal ejecta-CSM configuration. 
The symmetry axes of the two components are misaligned, so an asymmetric interaction region will be developed as the shock propagates through a nonuniform medium. It is intended to illustrate the geometric evolution of the interaction region, rather than produce an exact fit to the observed polarization.

The SN ejecta were initialized as a homologously expanding prolate
ellipsoid with a broken power-law density profile, 
\begin{equation}
 \rho_{\rm ej}(r_{\rm ej})=\frac{f_3M_{\rm ej}}
 {4\pi v_{\rm br}^3t^3}
 \begin{cases}
  (r_{\rm ej}/v_{\rm br}t)^{-\delta}, & r_{\rm ej}\leq v_{\rm br}t,\\
  (r_{\rm ej}/v_{\rm br}t)^{-m}, & r_{\rm ej}>v_{\rm br}t,
 \end{cases}
 \label{eq:ejecta_density}
\end{equation}
and
\begin{equation}
 f_l=\frac{(m-l)(l-\delta)}
 {m-\delta-(l-\delta)(v_{\rm br}/v_{\rm max})^{m-l}}.
 \label{eq:fl}
\end{equation}
Here $v_{\rm br}$ denotes the break velocity, and $v_{\rm max}$ is the maximum ejecta velocity. We adopt $f_{3}$ in Equation~(\ref{eq:ejecta_density}) to ensure the integrated mass equals $M_{\rm ej}$.
The simulations start at $t = t_{\rm 0} = 1.73 \times 10^4$~s. 
The ellipsoidal radius $r_{\rm ej}$ is defined as
\begin{equation}
r_{\rm ej}
=
\left(
\frac{x^2+y^2}{a^2}
+
\frac{z^2}{b^2}
\right)^{1/2},
\end{equation}
where $(x, y, z)$ are Cartesian coordinates in the frame of the ejecta, and $a$ and $c$ are the equatorial and polar semiaxes, respectively.
We adopt $\delta=1$, $m=10$, $M_{\rm ej}=1.85\,M_\odot$, and $E_{\rm kin}=10^{51}$~erg, which give $v_{\rm br}=8.81\times10^8~{\rm cm~s^{-1}}$. The ejecta axis ratio is $A_{\rm ej}=b/a=1.25$, and their symmetry axis
is tilted by \(\theta_{\rm ej}=+8^\circ\), measured counterclockwise from the \(x\)-axis in the \(x\)--\(z\) plane as viewed from the \(+y\) direction.

We describe the CSM using volume-preserving, oblate ellipsoidal
isodensity surfaces. Following the dense-shell CSM model adopted by \citet{deWet2026}, we assume a density structure of a dense inner shell with a steeply declining outer edge, surrounded by low-density $r^{-2}$ stellar wind. This can be expressed as 
\begin{equation}
 \rho_{\rm CSM}(r_{\rm CSM})=\rho_0
 \begin{cases}
 (r_{\rm CSM}/R_0)^{-s_1}, & R_0<r_{\rm CSM}\leq R_1,\\
 (R_1/R_0)^{-s_1}(r_{\rm CSM}/R_1)^{-s_2}, & R_1<r_{\rm CSM}\leq R_2,\\
 (R_1/R_0)^{-s_1}(R_2/R_1)^{-s_2}(r_{\rm CSM}/R_2)^{-s_3},
 & R_2<r_{\rm CSM}\leq R_{\rm CSM},
 \end{cases}
 \label{eq:csm_density}
\end{equation}
where $s_{1}$, $s_{2}$, and $s_{3}$ are the power-law slopes of the dense inner shell, the steep outer edge, and the stellar wind, respectively. The break radii are denoted by $R_{1}$ and $R_{2}$. We adopt $\rho_0=5.65\times10^{-13}\ {\rm g\,cm^{-3}}$,
$(s_1,s_2,s_3)=(0.5,30,2)$, and
$(R_1,R_2)=(1.037,1.5)\times10^{15}$ cm. 
The factors joining the three segments equal unity at the break radii, so Equation~(\ref{eq:csm_density}) is continuous at both breaks. 
The CSM axis ratio is $A_{\rm CSM}\equiv a/b=1.20$, where $a$ and $c$ are respectively the equatorial and polar semiaxes, and its symmetry axis is inclined by $\theta_{\rm CSM}=-25^\circ$ from the $x$ axis in the $x$--$z$ plane. The misalignment with the ejecta axis breaks axial symmetry and produces a direction-dependent ejecta-CSM interaction. 

We assume a hydrogen-free, helium-rich CSM composition with
$(X_{\rm H},X_{\rm He},Z)=(0,0.98,0.02)$, where $X_{\rm H}$,
$X_{\rm He}$, and $Z$ denote the mass fractions of hydrogen, helium, and metals, respectively. The opacity includes electron scattering and free-free absorption. The electron-scattering opacity is defined following~\citet{Rybicki1979}, 
\begin{equation}
 \kappa_{\rm es}=0.2(1+X_{\rm H})\chi_{\rm ion}
 \ {\rm cm^2\,g^{-1}}.
 \label{eq:kappa_es}
\end{equation}
where $\chi_{\rm ion}$ is the ionization factor defined below.
We adopt an electron-scattering opacity of \(\kappa_{\rm es}=0.2~\mathrm{cm^2\,g^{-1}}\), appropriate for fully ionized, helium-rich CSM. 
The free-free absorption opacity is
\begin{equation}
 \kappa_{\rm ff}=3.7\times10^{22}\,
 \chi_{\rm ion}(1+X_{\rm H})(X_{\rm H}+X_{\rm He})
 \rho T_{\rm g}^{-3.5}\ {\rm cm^2\,g^{-1}},
 \label{eq:kappa_ff}
\end{equation}
where $T_{\rm g}$ is the gas temperature \citep{Rybicki1979}. 
The factor $\chi_{\rm ion}$ describes the reduction of both opacities due to partial ionization. 
\citet{Faran2019} proposed that the ionization degree scales as \(\chi_{\rm ion}\propto T^\beta\), with \(\beta=11\).
We adopt a smooth recombination prescription that reproduces this scaling,
\begin{equation}
 \chi_{\rm ion}=
 \left[
 1+\left(\frac{T_{\rm g}}{T_{\rm rec}}\right)^{-\beta}
 \right]^{-1}.
 \label{eq:ionization}
\end{equation}
For the helium-rich CSM, we adopt a recombination temperature of \(T_{\rm rec}=1.2\times10^{4}\,\mathrm{K}\), which is higher than for hydrogen because of the high ionization potential of helium~\citep{Suzuki2020}.
Although this prescription simplifies helium recombination, it captures the resulting reduction in electron-scattering opacity. Our aim is to assess how this reduction affects the polarization of hydrogen-free material, rather than to model recombination in detail.

We map each 3D radiation-hydrodynamic snapshot onto a $100^{3}$ Cartesian MCPol transport grid.
At each epoch, we define an angle-dependent emission surface $r_{\rm s}(\theta,\phi)$, where $\theta$ and $\phi$ are the polar and azimuthal angles about the explosion center. 
While the forward shock is still embedded in the optically thick, dense CSM (its radius still lies inside the $\tau=1$ surface), we take the shock surface as the emitting surface. As the shock propagates outward, once the $\tau=1$ surfaces has receded within the shocked material, we use the $\tau=1$ surface as the emitting surface instead.
The local outward radiative flux on this surface, $F_r[r_{\rm s}(\theta,\phi),\theta,\phi]$, sets the luminosity assigned to each surface element, 
\begin{equation}
 dL(\theta,\phi)=
 F_r(\theta,\phi)
 r_{\rm s}^2(\theta,\phi)\,d\Omega .
 \label{eq:source_luminosity}
\end{equation}
Unpolarized photon packets are launched from the emission surface with a probability
\begin{equation}
 p(\theta,\phi)=\frac{dL(\theta,\phi)}
 {\int dL(\theta,\phi)}.
 \label{eq:packet_probability}
\end{equation}
This procedure preserves the intrinsically anisotropic surface-brightness distribution predicted by the radiation-hydrodynamic calculation. 
The packets are subsequently propagated through the mapped 3D medium, undergoing multiple electron scatterings until they escape.
We collect the escaped packets along the $+y$ viewing direction in 16 angular bins and measure the emergent Stokes parameters $(I,Q,U)$ in a reference frame defined by the projected $z$-axis.
The normalized Stokes parameters, $Q/I$ and $U/I$, provide the model continuum polarization at each epoch.

\subsection{The Effect of Nonuniform Luminance Distribution on Polarization}~\label{sec:appendix_contpol}

When the radiation surface has a nonuniform intensity distribution, continuous polarization suffers additional effects beyond those caused by the basic geometric factors. Therefore, we performed a series of continuous polarization simulations using MCPol to investigate the influence of this nonuniformity on the final results.
%We performed a suite of continuum polarization simulations with MCPol to constrain the allowed range of parameter space (e.g., the scattering geometries), following methods widely used in SN polarization studies \citep{Hoflich1991, Kasen2003, 2020ApJ...902...46Y, 2023MNRAS.520..560H}. 
For the misaligned ejecta-CSM interaction, we approximate the dominant electron-scattering region as an oblate ellipsoid during the CSM-dominated phase and as a prolate ellipsoid once the CDS becomes dominant. 
The density structure follows a power law in the ellipsoidal radius, $\rho\propto r^{-n}$. For the dense CSM, $n$ is expected to lie within $0$--$1$, and we adopt $n=0.5$ as the fiducial value. The strongly compressed CDS is instead characterized by a much steeper profile with $n\geq12$.
The asphericity of the oblate CSM and prolate CDS is parameterized by $A_{\rm oblate}=a/c$ and $A_{\rm prolate}=c/a$, respectively, where $a$ and $c$ are the equatorial and polar semiaxes. In both cases, $A=1$ corresponds to a sphere, and a larger $A$ indicates greater asphericity. We consider $1\leq A\leq1.3$.

\begin{figure*}[ht]  
    \centering
    \includegraphics[width=1.0\textwidth]{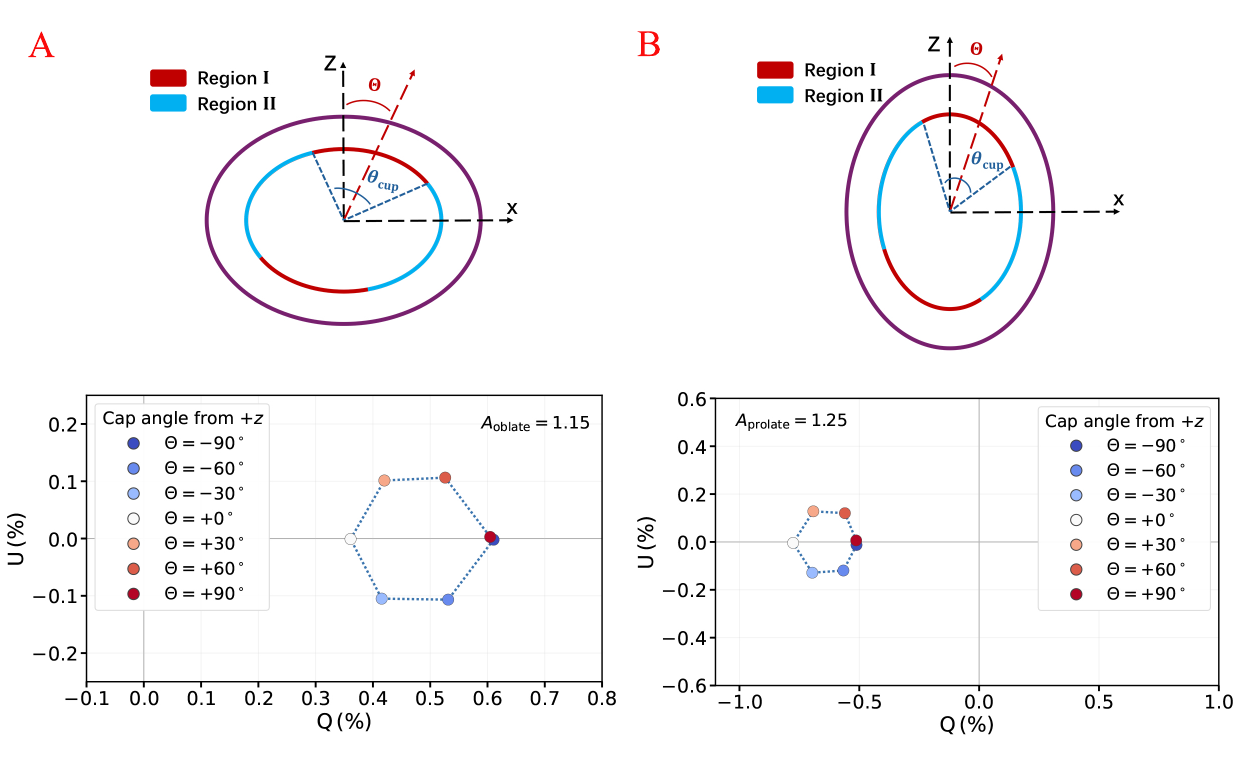}
    \vspace{-8.0mm}
     \captionsetup{name=\suppmattwo Figure }
    \caption{Effects of anisotropic illumination on the continuum polarization for the oblate (Panel~A) and prolate (Panel~B) models. 
    %Panels A and B show schematic representations of the oblate and prolate models (top) and their corresponding $Q$--$U$ distributions for different orientations $\theta$ (bottom), respectively. In both panels, Regions I and II are shown in red and blue, respectively; point colors indicate $\theta$, and dotted lines connect the results. 
    \textit{Top:} schematic of each model, with the cap-like Region~I in red and the remaining Region~II in blue. Here $\Theta$ denotes the inclination of the axis of Region~I relative to the $+z$-axis, and $\theta_{\rm cap}$ is the opening angle of the cap. 
    \textit{Bottom:} the corresponding $Q-U$ distributions for different $\Theta$ as shown by the color-coded points and indicated by the legend. Dotted lines of the same color connect the results for the given given axis ratio $A_{\rm oblate}$ and $A_{\rm prolate}$ as labeled in each subpanel.
    All models adopt $\theta_{\rm cup}=80^\circ$ and $F_{\rm I}/F_{\rm II}=0.8$. 
    Varying $\Theta$ produces an approximately ring-like locus around the uniform-illumination solution. The oblate and prolate cases represent the CSM- and CDS-dominated phases, respectively.}
~\label{fig:Appendix_mcpol} 
\end{figure*}

As the shock propagates outward through the dense CSM, the radiation intensity is relatively stronger along the high-density equatorial direction. We divide the emitting surface into Regions I and II using a cap with opening angle $\theta_{\rm cup}$, and assign radiative fluxes $F_1$ and $F_2$ to the two regions, respectively. For the early CSM dominated phase, we parameterize this anisotropy for an oblate geometry by setting $F_{\rm I}/F_{\rm II}=0.8$. The number of photon packets emitted from the two regions is then given by $N_{\rm I}/N_{\rm II}\approx(F_{\rm I}S_{\rm I})/(F_{\rm II}S_{\rm II})$,
where $S_{\rm I}$ and $S_{\rm II}$ are the surface areas of Regions~I and II, respectively.
For $\theta_{\rm cup}=80^\circ$, $F_{\rm I}/F_{\rm II}=0.8$, and
$A_{\rm out}=1.15$, varying the orientation of Region~I (i.e., $\Theta$) traces an approximately ring-like locus around the uniform-illumination solution in the $Q-U$ plane (Appendix Figure~\ref{fig:Appendix_mcpol}A). The individual Stokes $Q$ and $U$ components vary by up to $\sim \pm 0.1$\%.
The size of the locus is primarily determined primarily by the flux contrast $F_{\rm I}/F_{\rm II}$ and the relative surface areas of the two regions.

%Once the shock emerges from the dense CSM and the CDS becomes the dominant scattering region, we approximate its geometry with a prolate ellipsoid. Faster expansion along the lower-density direction can make the outer boundary of this region more elongated than the inner boundary, thereby enhancing the continuum polarization. Appendix Figure~\ref{fig:Appendix_mcpol}E considers a fixed inner boundary with $A_{\rm in}=1.15$, and varies the outer axis ratio from $A_{\rm out}=1.15$ to $1.2$ to mimic the effect of anisotropic expansion. The increasing asphericity of the outer boundary leads to a substantial increase in the continuum polarization. A change of only $\sim4$\% in the outer axis ratio is therefore sufficient to alter the polarization noticeably, which shows that the polarization at these epochs is sensitive to the shape of the outer boundary.

We also apply the same two-region prescription to account for anisotropic emission from the CDS. We assume that the weaker interaction and more rapid expansion along the lower-density direction produce a lower local radiative flux than along the higher-density direction.
For $\theta_{\rm cup}=80^\circ$, $F_{\rm I}/F_{\rm II}=0.8$, and $A_{\rm in}=1.25$, changing the orientation $\theta$ again produces a ring-like distribution around the uniform-illumination solution (Appendix Figure~\ref{fig:Appendix_mcpol}B). 
These tests indicate that, for sufficiently aspherical configurations, the continuum polarization is controlled primarily by the global scattering geometry.
Anisotropic illumination can significantly affect both the polarization degree and PA once its contribution becomes comparable to the geometric polarization.

%% For this sample we use BibTeX plus aasjournalv7.bst to generate the

%\clearpage
\bibliography{SN2025kzr}{}

\begin{thebibliography}{}
\expandafter\ifx\csname natexlab\endcsname\relax\def\natexlab#1{#1}\fi
\providecommand{\url}[1]{\href{#1}{#1}}
\providecommand{\dodoi}[1]{doi:~\href{http://doi.org/#1}{\nolinkurl{#1}}}
\providecommand{\doeprint}[1]{\href{http://ascl.net/#1}{\nolinkurl{http://ascl.net/#1}}}
\providecommand{\doarXiv}[1]{\href{https://arxiv.org/abs/#1}{\nolinkurl{https://arxiv.org/abs/#1}}}

% type= article
\bibitem[{J.~P. {Anderson} {et~al.}(2026){Anderson}, {Aster}, {Bulla}, {Chen}, {Fraser}, {Galbany}, {Guti{\'e}rrez}, {Inserra}, {Killestein}, {Leloudas}, {Lyman}, {Maeda}, {Maguire}, {Mason}, {Moriya}, {Pastorello}, {Taubenberger}, {Pursiainen}, \& {Wichern}}]{2026A&A...710A..35A}
{Anderson}, J.~P., {Aster}, C., {Bulla}, M., {et~al.} 2026, \bibinfo{title}{{Nested, asymmetric H─He circumstellar shells in the Type Icn/Ibn SN 2024abvb},} \aap, 710, A35, \dodoi{10.1051/0004-6361/202659241}

% type= article
\bibitem[{G.~C. {Anupama} {et~al.}(2009){Anupama}, {Sahu}, {Gurugubelli}, {Prabhu}, {Tominaga}, {Tanaka}, \& {Nomoto}}]{2009MNRAS.392..894A}
{Anupama}, G.~C., {Sahu}, D.~K., {Gurugubelli}, U.~K., {et~al.} 2009, \bibinfo{title}{{Optical photometry and spectroscopy of the Type Ibn supernova SN 2006jc until the onset of dust formation},} \mnras, 392, 894, \dodoi{10.1111/j.1365-2966.2008.14129.x}

% type= article
\bibitem[{I. {Appenzeller} {et~al.}(1998){Appenzeller}, {Fricke}, {F{\"u}rtig}, {G{\"a}ssler}, {H{\"a}fner}, {Harke}, {Hess}, {Hummel}, {J{\"u}rgens}, {Kudritzki}, {Mantel}, {Meisl}, {Muschielok}, {Nicklas}, {Rupprecht}, {Seifert}, {Stahl}, {Szeifert}, \& {Tarantik}}]{1998Msngr..94....1A}
{Appenzeller}, I., {Fricke}, K., {F{\"u}rtig}, W., {et~al.} 1998, \bibinfo{title}{{Successful commissioning of FORS1 - the first optical instrument on the VLT.},} The Messenger, 94, 1

% type= article
\bibitem[{ {Astropy Collaboration} {et~al.}(2013){Astropy Collaboration}, {Robitaille}, {Tollerud}, {Greenfield}, {Droettboom}, {Bray}, {Aldcroft}, {Davis}, {Ginsburg}, {Price-Whelan}, {Kerzendorf}, {Conley}, {Crighton}, {Barbary}, {Muna}, {Ferguson}, {Grollier}, {Parikh}, {Nair}, {Unther}, {Deil}, {Woillez}, {Conseil}, {Kramer}, {Turner}, {Singer}, {Fox}, {Weaver}, {Zabalza}, {Edwards}, {Azalee Bostroem}, {Burke}, {Casey}, {Crawford}, {Dencheva}, {Ely}, {Jenness}, {Labrie}, {Lim}, {Pierfederici}, {Pontzen}, {Ptak}, {Refsdal}, {Servillat}, \& {Streicher}}]{2013A&A...558A..33A}
{Astropy Collaboration}, {Robitaille}, T.~P., {Tollerud}, E.~J., {et~al.} 2013, \bibinfo{title}{{Astropy: A community Python package for astronomy},} \aap, 558, A33, \dodoi{10.1051/0004-6361/201322068}

% type= article
\bibitem[{ {Astropy Collaboration} {et~al.}(2018){Astropy Collaboration}, {Price-Whelan}, {Sip{\H{o}}cz}, {G{\"u}nther}, {Lim}, {Crawford}, {Conseil}, {Shupe}, {Craig}, {Dencheva}, {Ginsburg}, {VanderPlas}, {Bradley}, {P{\'e}rez-Su{\'a}rez}, {de Val-Borro}, {Aldcroft}, {Cruz}, {Robitaille}, {Tollerud}, {Ardelean}, {Babej}, {Bach}, {Bachetti}, {Bakanov}, {Bamford}, {Barentsen}, {Barmby}, {Baumbach}, {Berry}, {Biscani}, {Boquien}, {Bostroem}, {Bouma}, {Brammer}, {Bray}, {Breytenbach}, {Buddelmeijer}, {Burke}, {Calderone}, {Cano Rodr{\'\i}guez}, {Cara}, {Cardoso}, {Cheedella}, {Copin}, {Corrales}, {Crichton}, {D'Avella}, {Deil}, {Depagne}, {Dietrich}, {Donath}, {Droettboom}, {Earl}, {Erben}, {Fabbro}, {Ferreira}, {Finethy}, {Fox}, {Garrison}, {Gibbons}, {Goldstein}, {Gommers}, {Greco}, {Greenfield}, {Groener}, {Grollier}, {Hagen}, {Hirst}, {Homeier}, {Horton}, {Hosseinzadeh}, {Hu}, {Hunkeler}, {Ivezi{\'c}}, {Jain}, {Jenness}, {Kanarek}, {Kendrew}, {Kern}, {Kerzendorf}, {Khvalko}, {King}, {Kirkby}, {Kulkarni},
  {Kumar}, {Lee}, {Lenz}, {Littlefair}, {Ma}, {Macleod}, {Mastropietro}, {McCully}, {Montagnac}, {Morris}, {Mueller}, {Mumford}, {Muna}, {Murphy}, {Nelson}, {Nguyen}, {Ninan}, {N{\"o}the}, {Ogaz}, {Oh}, {Parejko}, {Parley}, {Pascual}, {Patil}, {Patil}, {Plunkett}, {Prochaska}, {Rastogi}, {Reddy Janga}, {Sabater}, {Sakurikar}, {Seifert}, {Sherbert}, {Sherwood-Taylor}, {Shih}, {Sick}, {Silbiger}, {Singanamalla}, {Singer}, {Sladen}, {Sooley}, {Sornarajah}, {Streicher}, {Teuben}, {Thomas}, {Tremblay}, {Turner}, {Terr{\'o}n}, {van Kerkwijk}, {de la Vega}, {Watkins}, {Weaver}, {Whitmore}, {Woillez}, {Zabalza}, \& {Astropy Contributors}}]{2018AJ....156..123A}
{Astropy Collaboration}, {Price-Whelan}, A.~M., {Sip{\H{o}}cz}, B.~M., {et~al.} 2018, \bibinfo{title}{{The Astropy Project: Building an Open-science Project and Status of the v2.0 Core Package},} \aj, 156, 123, \dodoi{10.3847/1538-3881/aabc4f}

% type= article
\bibitem[{ {Astropy Collaboration} {et~al.}(2022){Astropy Collaboration}, {Price-Whelan}, {Lim}, {Earl}, {Starkman}, {Bradley}, {Shupe}, {Patil}, {Corrales}, {Brasseur}, {N{\"o}the}, {Donath}, {Tollerud}, {Morris}, {Ginsburg}, {Vaher}, {Weaver}, {Tocknell}, {Jamieson}, {van Kerkwijk}, {Robitaille}, {Merry}, {Bachetti}, {G{\"u}nther}, {Aldcroft}, {Alvarado-Montes}, {Archibald}, {B{\'o}di}, {Bapat}, {Barentsen}, {Baz{\'a}n}, {Biswas}, {Boquien}, {Burke}, {Cara}, {Cara}, {Conroy}, {Conseil}, {Craig}, {Cross}, {Cruz}, {D'Eugenio}, {Dencheva}, {Devillepoix}, {Dietrich}, {Eigenbrot}, {Erben}, {Ferreira}, {Foreman-Mackey}, {Fox}, {Freij}, {Garg}, {Geda}, {Glattly}, {Gondhalekar}, {Gordon}, {Grant}, {Greenfield}, {Groener}, {Guest}, {Gurovich}, {Handberg}, {Hart}, {Hatfield-Dodds}, {Homeier}, {Hosseinzadeh}, {Jenness}, {Jones}, {Joseph}, {Kalmbach}, {Karamehmetoglu}, {Ka{\l}uszy{\'n}ski}, {Kelley}, {Kern}, {Kerzendorf}, {Koch}, {Kulumani}, {Lee}, {Ly}, {Ma}, {MacBride}, {Maljaars}, {Muna}, {Murphy}, {Norman},
  {O'Steen}, {Oman}, {Pacifici}, {Pascual}, {Pascual-Granado}, {Patil}, {Perren}, {Pickering}, {Rastogi}, {Roulston}, {Ryan}, {Rykoff}, {Sabater}, {Sakurikar}, {Salgado}, {Sanghi}, {Saunders}, {Savchenko}, {Schwardt}, {Seifert-Eckert}, {Shih}, {Jain}, {Shukla}, {Sick}, {Simpson}, {Singanamalla}, {Singer}, {Singhal}, {Sinha}, {Sip{\H{o}}cz}, {Spitler}, {Stansby}, {Streicher}, {{\v{S}}umak}, {Swinbank}, {Taranu}, {Tewary}, {Tremblay}, {de Val-Borro}, {Van Kooten}, {Vasovi{\'c}}, {Verma}, {de Miranda Cardoso}, {Williams}, {Wilson}, {Winkel}, {Wood-Vasey}, {Xue}, {Yoachim}, {Zhang}, {Zonca}, \& {Astropy Project Contributors}}]{2022ApJ...935..167A}
{Astropy Collaboration}, {Price-Whelan}, A.~M., {Lim}, P.~L., {et~al.} 2022, \bibinfo{title}{{The Astropy Project: Sustaining and Growing a Community-oriented Open-source Project and the Latest Major Release (v5.0) of the Core Package},} \apj, 935, 167, \dodoi{10.3847/1538-4357/ac7c74}

% type= article
\bibitem[{C. {Bilinski} {et~al.}(2024){Bilinski}, {Smith}, {Williams}, {Smith}, {Leonard}, {Hoffman}, {Andrews}, \& {Milne}}]{2024MNRAS.529.1104B}
{Bilinski}, C., {Smith}, N., {Williams}, G.~G., {et~al.} 2024, \bibinfo{title}{{Multi-epoch spectropolarimetry for a sample of Type IIn Supernovae: persistent asymmetry in dusty circumstellar material},} \mnras, 529, 1104, \dodoi{10.1093/mnras/stae380}

% type= article
\bibitem[{J.~M. {Blondin} {et~al.}(2003){Blondin}, {Mezzacappa}, \& {DeMarino}}]{2003ApJ...584..971B}
{Blondin}, J.~M., {Mezzacappa}, A., \& {DeMarino}, C. 2003, \bibinfo{title}{{Stability of Standing Accretion Shocks, with an Eye toward Core-Collapse Supernovae},} \apj, 584, 971, \dodoi{10.1086/345812}

% type= article
\bibitem[{A. {Burrows} {et~al.}(2007){Burrows}, {Dessart}, {Livne}, {Ott}, \& {Murphy}}]{2007ApJ...664..416B}
{Burrows}, A., {Dessart}, L., {Livne}, E., {Ott}, C.~D., \& {Murphy}, J. 2007, \bibinfo{title}{{Simulations of Magnetically Driven Supernova and Hypernova Explosions in the Context of Rapid Rotation},} \apj, 664, 416, \dodoi{10.1086/519161}

% type= article
\bibitem[{A. {Cikota} {et~al.}(2017){Cikota}, {Patat}, {Cikota}, \& {Faran}}]{2017MNRAS.464.4146C}
{Cikota}, A., {Patat}, F., {Cikota}, S., \& {Faran}, T. 2017, \bibinfo{title}{{Linear spectropolarimetry of polarimetric standard stars with VLT/FORS2},} \mnras, 464, 4146, \dodoi{10.1093/mnras/stw2545}

% type= article
\bibitem[{S. {de Wet} {et~al.}(2026){de Wet}, {Leloudas}, {Buckley}, {Erasmus}, {Groot}, {Zimmerman}, {Chen}, {Tampo}, {Pursiainen}, {Killestein}, {Stoppa}, {Jaisawal}, {Gal-Yam}, {Maeda}, {Anderson}, {Chen}, {Gromadzki}, {Guti{\'e}rrez}, {Kankare}, {M{\"u}ller-Bravo}, {Pessi}, {Smartt}, {Sollerman}, {Tartaglia}, {Young}, {Alarcon}, {Smith}, {Stevance}, {de Boer}, {Chambers}, {Lin}, {Lowe}, {Minguez}, {Nicholl}, {Paek}, \& {Wainscoat}}]{deWet2026}
{de Wet}, S., {Leloudas}, G., {Buckley}, D.~A.~H., {et~al.} 2026, \bibinfo{title}{{The flash-ionised SN Ibn 2025kzr: H-free CSM formed during a precursor outburst 55 days prior to explosion},} arXiv e-prints, arXiv:2606.18962, \dodoi{10.48550/arXiv.2606.18962}

% type= article
\bibitem[{L. {Dessart} {et~al.}(2022){Dessart}, {Hillier}, \& {Kuncarayakti}}]{2022A&A...658A.130D}
{Dessart}, L., {Hillier}, D.~J., \& {Kuncarayakti}, H. 2022, \bibinfo{title}{{Helium stars exploding in circumstellar material and the origin of Type Ibn supernovae},} \aap, 658, A130, \dodoi{10.1051/0004-6361/202142436}

% type= article
\bibitem[{Y. {Dong} {et~al.}(2024){Dong}, {Tsuna}, {Valenti}, {Sand}, {Andrews}, {Bostroem}, {Hosseinzadeh}, {Hoang}, {Jha}, {Janzen}, {Jencson}, {Lundquist}, {Mehta}, {Ravi}, {Meza Retamal}, {Pearson}, {Shrestha}, {Bonanos}, {Howell}, {Smith}, {Farah}, {Hiramatsu}, {Itagaki}, {McCully}, {Newsome}, {Padilla Gonzalez}, {Paraskeva}, {Pellegrino}, {Terreran}, {Haislip}, {Kouprianov}, \& {Reichart}}]{2024ApJ...977..254D}
{Dong}, Y., {Tsuna}, D., {Valenti}, S., {et~al.} 2024, \bibinfo{title}{{SN2023fyq: A Type Ibn Supernova with Long-standing Precursor Activity Due to Binary Interaction},} \apj, 977, 254, \dodoi{10.3847/1538-4357/ad8de6}

% type= article
\bibitem[{A. {Ercolino} {et~al.}(2025){Ercolino}, {Jin}, {Langer}, \& {Dessart}}]{2025A&A...696A.103E}
{Ercolino}, A., {Jin}, H., {Langer}, N., \& {Dessart}, L. 2025, \bibinfo{title}{{Mass-transferring binary stars as progenitors of interacting hydrogen-free supernovae},} \aap, 696, A103, \dodoi{10.1051/0004-6361/202453426}

% type= article
\bibitem[{T. {Faran} {et~al.}(2019){Faran}, {Goldfriend}, {Nakar}, \& {Sari}}]{Faran2019}
{Faran}, T., {Goldfriend}, T., {Nakar}, E., \& {Sari}, R. 2019, \bibinfo{title}{{Recombination Effects on Supernova Light Curves},} \apj, 879, 20, \dodoi{10.3847/1538-4357/ab218a}

% type= article
\bibitem[{R.~J. {Foley} {et~al.}(2007){Foley}, {Smith}, {Ganeshalingam}, {Li}, {Chornock}, \& {Filippenko}}]{2007ApJ...657L.105F}
{Foley}, R.~J., {Smith}, N., {Ganeshalingam}, M., {et~al.} 2007, \bibinfo{title}{{SN 2006jc: A Wolf-Rayet Star Exploding in a Dense He-rich Circumstellar Medium},} \apjl, 657, L105, \dodoi{10.1086/513145}

% type= inproceedings
\bibitem[{L. {Fossati} {et~al.}(2007){Fossati}, {Bagnulo}, {Mason}, \& {Landi Degl'Innocenti}}]{2007ASPC..364..503F}
{Fossati}, L., {Bagnulo}, S., {Mason}, E., \& {Landi Degl'Innocenti}, E. 2007, \bibinfo{title}{{Standard Stars for Linear Polarization Observed with FORS1},} in Astronomical Society of the Pacific Conference Series, Vol. 364, The Future of Photometric, Spectrophotometric and Polarimetric Standardization, ed. C.~{Sterken}, 503

% type= article
\bibitem[{J. {Fuller}(2017){Fuller}}]{Fuller2017}
{Fuller}, J. 2017, \bibinfo{title}{{Pre-supernova outbursts via wave heating in massive stars - I. Red supergiants},} \mnras, 470, 1642, \dodoi{10.1093/mnras/stx1314}

% type= article
\bibitem[{J. {Fuller} \& S. {Ro}(2018){Fuller} \& {Ro}}]{Fuller2018}
{Fuller}, J., \& {Ro}, S. 2018, \bibinfo{title}{{Pre-supernova outbursts via wave heating in massive stars - II. Hydrogen-poor stars},} \mnras, 476, 1853, \dodoi{10.1093/mnras/sty369}

% type= incollection
\bibitem[{A. {Gal-Yam}(2017){Gal-Yam}}]{2017hsn..book..195G}
{Gal-Yam}, A. 2017, \bibinfo{title}{{Observational and Physical Classification of Supernovae},} in Handbook of Supernovae, ed. A.~W. {Alsabti} \& P.~{Murdin} (Springer International Publishing), 195, \dodoi{10.1007/978-3-319-21846-5_35}

% type= article
\bibitem[{C.~R. {Harris} {et~al.}(2020){Harris}, {Millman}, {van der Walt}, {Gommers}, {Virtanen}, {Cournapeau}, {Wieser}, {Taylor}, {Berg}, {Smith}, {Kern}, {Picus}, {Hoyer}, {van Kerkwijk}, {Brett}, {Haldane}, {del R{\'\i}o}, {Wiebe}, {Peterson}, {G{\'e}rard-Marchant}, {Sheppard}, {Reddy}, {Weckesser}, {Abbasi}, {Gohlke}, \& {Oliphant}}]{2020Natur.585..357H}
{Harris}, C.~R., {Millman}, K.~J., {van der Walt}, S.~J., {et~al.} 2020, \bibinfo{title}{{Array programming with NumPy},} \nat, 585, 357, \dodoi{10.1038/s41586-020-2649-2}

% type= article
\bibitem[{P. {Hoeflich} {et~al.}(2023){Hoeflich}, {Yang}, {Baade}, {Cikota}, {Maund}, {Mishra}, {Patat}, {Patra}, {Wang}, {Wheeler}, {Filippenko}, {Gal-Yam}, \& {Schulze}}]{2023MNRAS.520..560H}
{Hoeflich}, P., {Yang}, Y., {Baade}, D., {et~al.} 2023, \bibinfo{title}{{The core normal Type Ia supernova 2019np - an overall spherical explosion with an aspherical surface layer and an aspherical $^{56}$Ni core},} \mnras, 520, 560, \dodoi{10.1093/mnras/stad172}

% type= article
\bibitem[{J.~L. {Hoffman} {et~al.}(2008){Hoffman}, {Leonard}, {Chornock}, {Filippenko}, {Barth}, \& {Matheson}}]{2008ApJ...688.1186H}
{Hoffman}, J.~L., {Leonard}, D.~C., {Chornock}, R., {et~al.} 2008, \bibinfo{title}{{The Dual-Axis Circumstellar Environment of the Type IIn Supernova 1997eg},} \apj, 688, 1186, \dodoi{10.1086/592261}

% type= article
\bibitem[{P. {Hoflich}(1991){Hoflich}}]{Hoflich1991}
{Hoflich}, P. 1991, \bibinfo{title}{{Asphericity effects in scatterring dominated photospheres.},} \aap, 246, 481

% type= article
\bibitem[{G. {Hosseinzadeh} {et~al.}(2017){Hosseinzadeh}, {Arcavi}, {Valenti}, {McCully}, {Howell}, {Johansson}, {Sollerman}, {Pastorello}, {Benetti}, {Cao}, {Cenko}, {Clubb}, {Corsi}, {Duggan}, {Elias-Rosa}, {Filippenko}, {Fox}, {Fremling}, {Horesh}, {Karamehmetoglu}, {Kasliwal}, {Marion}, {Ofek}, {Sand}, {Taddia}, {Zheng}, {Fraser}, {Gal-Yam}, {Inserra}, {Laher}, {Masci}, {Rebbapragada}, {Smartt}, {Smith}, {Sullivan}, {Surace}, \& {Wo{\'z}niak}}]{2017ApJ...836..158H}
{Hosseinzadeh}, G., {Arcavi}, I., {Valenti}, S., {et~al.} 2017, \bibinfo{title}{{Type Ibn Supernovae Show Photometric Homogeneity and Spectral Diversity at Maximum Light},} \apj, 836, 158, \dodoi{10.3847/1538-4357/836/2/158}

% type= article
\bibitem[{D.~A. {Howell} {et~al.}(2001){Howell}, {H{\"o}flich}, {Wang}, \& {Wheeler}}]{2001ApJ...556..302H}
{Howell}, D.~A., {H{\"o}flich}, P., {Wang}, L., \& {Wheeler}, J.~C. 2001, \bibinfo{title}{{Evidence for Asphericity in a Subluminous Type Ia Supernova: Spectropolarimetry of SN 1999by},} \apj, 556, 302, \dodoi{10.1086/321584}

% type= article
\bibitem[{J.~D. {Hunter}(2007){Hunter}}]{2007CSE.....9...90H}
{Hunter}, J.~D. 2007, \bibinfo{title}{{Matplotlib: A 2D Graphics Environment},} Computing in Science and Engineering, 9, 90, \dodoi{10.1109/MCSE.2007.55}

% type= article
\bibitem[{Y.-F. {Jiang}(2021){Jiang}}]{2021ApJS..253...49J}
{Jiang}, Y.-F. 2021, \bibinfo{title}{{An Implicit Finite Volume Scheme to Solve the Time-dependent Radiation Transport Equation Based on Discrete Ordinates},} \apjs, 253, 49, \dodoi{10.3847/1538-4365/abe303}

% type= article
\bibitem[{D. {Kasen} {et~al.}(2003){Kasen}, {Nugent}, {Wang}, {Howell}, {Wheeler}, {H{\"o}flich}, {Baade}, {Baron}, \& {Hauschildt}}]{Kasen2003}
{Kasen}, D., {Nugent}, P., {Wang}, L., {et~al.} 2003, \bibinfo{title}{{Analysis of the Flux and Polarization Spectra of the Type Ia Supernova SN 2001el: Exploring the Geometry of the High-Velocity Ejecta},} \apj, 593, 788, \dodoi{10.1086/376601}

% type= article
\bibitem[{A.~M. {Khokhlov} {et~al.}(1999){Khokhlov}, {H{\"o}flich}, {Oran}, {Wheeler}, {Wang}, \& {Chtchelkanova}}]{1999ApJ...524L.107K}
{Khokhlov}, A.~M., {H{\"o}flich}, P.~A., {Oran}, E.~S., {et~al.} 1999, \bibinfo{title}{{Jet-induced Explosions of Core Collapse Supernovae},} \apjl, 524, L107, \dodoi{10.1086/312305}

% type= article
\bibitem[{W. {Lu} {et~al.}(2023){Lu}, {Fuller}, {Quataert}, \& {Bonnerot}}]{2023MNRAS.519.1409L}
{Lu}, W., {Fuller}, J., {Quataert}, E., \& {Bonnerot}, C. 2023, \bibinfo{title}{{On rapid binary mass transfer - I. Physical model},} \mnras, 519, 1409, \dodoi{10.1093/mnras/stac3621}

% type= article
\bibitem[{K. {Maeda} \& T.~J. {Moriya}(2022){Maeda} \& {Moriya}}]{2022ApJ...927...25M}
{Maeda}, K., \& {Moriya}, T.~J. 2022, \bibinfo{title}{{Properties of Type Ibn Supernovae: Implications for the Progenitor Evolution and the Origin of a Population of Rapid Transients},} \apj, 927, 25, \dodoi{10.3847/1538-4357/ac4672}

% type= article
\bibitem[{T. {Matheson} {et~al.}(2000){Matheson}, {Filippenko}, {Chornock}, {Leonard}, \& {Li}}]{2000AJ....119.2303M}
{Matheson}, T., {Filippenko}, A.~V., {Chornock}, R., {Leonard}, D.~C., \& {Li}, W. 2000, \bibinfo{title}{{Helium Emission Lines in the Type IC Supernova 1999CQ},} \aj, 119, 2303, \dodoi{10.1086/301352}

% type= article
\bibitem[{J.~C. {Mauerhan} {et~al.}(2024){Mauerhan}, {Smith}, {Williams}, {Smith}, {Filippenko}, {Bilinski}, {Zheng}, {Brink}, {Hoffman}, {Leonard}, {Milne}, {Jeffers}, {Modak}, {Stegman}, \& {Zhang}}]{2024MNRAS.527.6090M}
{Mauerhan}, J.~C., {Smith}, N., {Williams}, G.~G., {et~al.} 2024, \bibinfo{title}{{Record-breaking polarization from the interacting superluminous supernova 2017hcc},} \mnras, 527, 6090, \dodoi{10.1093/mnras/stad3579}

% type= article
\bibitem[{J.~R. {Maund} {et~al.}(2007){Maund}, {Wheeler}, {Patat}, {Baade}, {Wang}, \& {H{\"o}flich}}]{2007MNRAS.381..201M}
{Maund}, J.~R., {Wheeler}, J.~C., {Patat}, F., {et~al.} 2007, \bibinfo{title}{{Spectropolarimetry of the Type Ib/c SN 2005bf},} \mnras, 381, 201, \dodoi{10.1111/j.1365-2966.2007.12230.x}

% type= article
\bibitem[{J.~R. {Maund} {et~al.}(2010){Maund}, {Wheeler}, {Wang}, {Baade}, {Clocchiatti}, {Patat}, {H{\"o}flich}, {Quinn}, \& {Zelaya}}]{2010ApJ...722.1162M}
{Maund}, J.~R., {Wheeler}, J.~C., {Wang}, L., {et~al.} 2010, \bibinfo{title}{{A Spectropolarimetric View on the Nature of the Peculiar Type I SN 2005hk},} \apj, 722, 1162, \dodoi{10.1088/0004-637X/722/2/1162}

% type= article
\bibitem[{B.~D. {Metzger}(2022){Metzger}}]{Metzger2022}
{Metzger}, B.~D. 2022, \bibinfo{title}{{Luminous Fast Blue Optical Transients and Type Ibn/Icn SNe from Wolf-Rayet/Black Hole Mergers},} \apj, 932, 84, \dodoi{10.3847/1538-4357/ac6d59}

% type= article
\bibitem[{B. {M{\"u}ller}(2020){M{\"u}ller}}]{2020LRCA....6....3M}
{M{\"u}ller}, B. 2020, \bibinfo{title}{{Hydrodynamics of core-collapse supernovae and their progenitors},} Living Reviews in Computational Astrophysics, 6, 3, \dodoi{10.1007/s41115-020-0008-5}

% type= article
\bibitem[{A. {Pastorello} {et~al.}(2007){Pastorello}, {Smartt}, {Mattila}, {Eldridge}, {Young}, {Itagaki}, {Yamaoka}, {Navasardyan}, {Valenti}, {Patat}, {Agnoletto}, {Augusteijn}, {Benetti}, {Cappellaro}, {Boles}, {Bonnet-Bidaud}, {Botticella}, {Bufano}, {Cao}, {Deng}, {Dennefeld}, {Elias-Rosa}, {Harutyunyan}, {Keenan}, {Iijima}, {Lorenzi}, {Mazzali}, {Meng}, {Nakano}, {Nielsen}, {Smoker}, {Stanishev}, {Turatto}, {Xu}, \& {Zampieri}}]{2007Natur.447..829P}
{Pastorello}, A., {Smartt}, S.~J., {Mattila}, S., {et~al.} 2007, \bibinfo{title}{{A giant outburst two years before the core-collapse of a massive star},} \nat, 447, 829, \dodoi{10.1038/nature05825}

% type= article
\bibitem[{A. {Pastorello} {et~al.}(2008){Pastorello}, {Mattila}, {Zampieri}, {Della Valle}, {Smartt}, {Valenti}, {Agnoletto}, {Benetti}, {Benn}, {Branch}, {Cappellaro}, {Dennefeld}, {Eldridge}, {Gal-Yam}, {Harutyunyan}, {Hunter}, {Kjeldsen}, {Lipkin}, {Mazzali}, {Milne}, {Navasardyan}, {Ofek}, {Pian}, {Shemmer}, {Spiro}, {Stathakis}, {Taubenberger}, {Turatto}, \& {Yamaoka}}]{2008MNRAS.389..113P}
{Pastorello}, A., {Mattila}, S., {Zampieri}, L., {et~al.} 2008, \bibinfo{title}{{Massive stars exploding in a He-rich circumstellar medium - I. Type Ibn (SN 2006jc-like) events},} \mnras, 389, 113, \dodoi{10.1111/j.1365-2966.2008.13602.x}

% type= article
\bibitem[{F. {Patat} {et~al.}(2009){Patat}, {Baade}, {H{\"o}flich}, {Maund}, {Wang}, \& {Wheeler}}]{2009A&A...508..229P}
{Patat}, F., {Baade}, D., {H{\"o}flich}, P., {et~al.} 2009, \bibinfo{title}{{VLT spectropolarimetry of the fast expanding type Ia SN 2006X},} \aap, 508, 229, \dodoi{10.1051/0004-6361/200810651}

% type= article
\bibitem[{F. {Patat} {et~al.}(2012){Patat}, {H{\"o}flich}, {Baade}, {Maund}, {Wang}, \& {Wheeler}}]{2012A&A...545A...7P}
{Patat}, F., {H{\"o}flich}, P., {Baade}, D., {et~al.} 2012, \bibinfo{title}{{VLT Spectropolarimetry of the Type Ia SN 2005ke. A step towards understanding subluminous events},} \aap, 545, A7, \dodoi{10.1051/0004-6361/201219146}

% type= article
\bibitem[{F. {Patat} {et~al.}(2010){Patat}, {Maund}, {Benetti}, {Botticella}, {Cappellaro}, {Harutyunyan}, \& {Turatto}}]{2010A&A...510A.108P}
{Patat}, F., {Maund}, J.~R., {Benetti}, S., {et~al.} 2010, \bibinfo{title}{{VLT spectropolarimetry of the optical transient in NGC 300. Evidence of asymmetry in the circumstellar dust},} \aap, 510, A108, \dodoi{10.1051/0004-6361/200913083}

% type= article
\bibitem[{F. {Patat} \& M. {Romaniello}(2006){Patat} \& {Romaniello}}]{2006PASP..118..146P}
{Patat}, F., \& {Romaniello}, M. 2006, \bibinfo{title}{{Error Analysis for Dual-Beam Optical Linear Polarimetry},} \pasp, 118, 146, \dodoi{10.1086/497581}

% type= article
\bibitem[{F. {Patat} {et~al.}(2011){Patat}, {Taubenberger}, {Benetti}, {Pastorello}, \& {Harutyunyan}}]{2011A&A...527L...6P}
{Patat}, F., {Taubenberger}, S., {Benetti}, S., {Pastorello}, A., \& {Harutyunyan}, A. 2011, \bibinfo{title}{{Asymmetries in the type IIn SN 2010jl},} \aap, 527, L6, \dodoi{10.1051/0004-6361/201016217}

% type= article
\bibitem[{C. {Pellegrino} {et~al.}(2022){Pellegrino}, {Howell}, {Vink{\'o}}, {Gangopadhyay}, {Xiang}, {Arcavi}, {Brown}, {Burke}, {Hiramatsu}, {Hosseinzadeh}, {Li}, {McCully}, {Misra}, {Newsome}, {Gonzalez}, {Pritchard}, {Valenti}, {Wang}, \& {Zhang}}]{2022ApJ...926..125P}
{Pellegrino}, C., {Howell}, D.~A., {Vink{\'o}}, J., {et~al.} 2022, \bibinfo{title}{{Circumstellar Interaction Powers the Light Curves of Luminous Rapidly Evolving Optical Transients},} \apj, 926, 125, \dodoi{10.3847/1538-4357/ac3e63}

% type= article
\bibitem[{D.~A. {Perley} {et~al.}(2020){Perley}, {Fremling}, {Sollerman}, {Miller}, {Dahiwale}, {Sharma}, {Bellm}, {Biswas}, {Brink}, {Bruch}, {De}, {Dekany}, {Drake}, {Duev}, {Filippenko}, {Gal-Yam}, {Goobar}, {Graham}, {Graham}, {Ho}, {Irani}, {Kasliwal}, {Kim}, {Kulkarni}, {Mahabal}, {Masci}, {Modak}, {Neill}, {Nordin}, {Riddle}, {Soumagnac}, {Strotjohann}, {Schulze}, {Taggart}, {Tzanidakis}, {Walters}, \& {Yan}}]{2020ApJ...904...35P}
{Perley}, D.~A., {Fremling}, C., {Sollerman}, J., {et~al.} 2020, \bibinfo{title}{{The Zwicky Transient Facility Bright Transient Survey. II. A Public Statistical Sample for Exploring Supernova Demographics},} \apj, 904, 35, \dodoi{10.3847/1538-4357/abbd98}

% type= article
\bibitem[{M. {Pursiainen} {et~al.}(2023){Pursiainen}, {Leloudas}, {Schulze}, {Charalampopoulos}, {Angus}, {Anderson}, {Bauer}, {Chen}, {Galbany}, {Gromadzki}, {Guti{\'e}rrez}, {Inserra}, {Lyman}, {M{\"u}ller-Bravo}, {Nicholl}, {Smartt}, {Tartaglia}, {Wiseman}, \& {Young}}]{Pursiainen2023}
{Pursiainen}, M., {Leloudas}, G., {Schulze}, S., {et~al.} 2023, \bibinfo{title}{{SN 2023emq: A Flash-ionized Ibn Supernova with Possible C III Emission},} \apjl, 959, L10, \dodoi{10.3847/2041-8213/ad103d}

% type= article
\bibitem[{E. {Quataert} {et~al.}(2016){Quataert}, {Fern{\'a}ndez}, {Kasen}, {Klion}, \& {Paxton}}]{Quataert2016}
{Quataert}, E., {Fern{\'a}ndez}, R., {Kasen}, D., {Klion}, H., \& {Paxton}, B. 2016, \bibinfo{title}{{Super-Eddington stellar winds driven by near-surface energy deposition},} \mnras, 458, 1214, \dodoi{10.1093/mnras/stw365}

% type= article
\bibitem[{E. {Quataert} \& J. {Shiode}(2012){Quataert} \& {Shiode}}]{2012MNRAS.423L..92Q}
{Quataert}, E., \& {Shiode}, J. 2012, \bibinfo{title}{{Wave-driven mass loss in the last year of stellar evolution: setting the stage for the most luminous core-collapse supernovae},} \mnras, 423, L92, \dodoi{10.1111/j.1745-3933.2012.01264.x}

% type= book
\bibitem[{G.~B. Rybicki \& A.~P. Lightman(1979)Rybicki \& Lightman}]{Rybicki1979}
Rybicki, G.~B., \& Lightman, A.~P. 1979, Radiative Processes in Astrophysics, Wiley-Interscience (New York: John Wiley \& Sons), xv+382

% type=
\bibitem[{ {Science Software Branch at STScI}(2012){Science Software Branch at STScI}}]{2012ascl.soft07011S}
{Science Software Branch at STScI}. 2012, {PyRAF: Python alternative for IRAF},, Astrophysics Source Code Library, record ascl:1207.011 \doeprint{1207.011}

% type= article
\bibitem[{J.~H. {Shiode} \& E. {Quataert}(2014){Shiode} \& {Quataert}}]{2014ApJ...780...96S}
{Shiode}, J.~H., \& {Quataert}, E. 2014, \bibinfo{title}{{Setting the Stage for Circumstellar Interaction in Core-Collapse Supernovae. II. Wave-driven Mass Loss in Supernova Progenitors},} \apj, 780, 96, \dodoi{10.1088/0004-637X/780/1/96}

% type= article
\bibitem[{I. {Shivvers} {et~al.}(2016){Shivvers}, {Zheng}, {Mauerhan}, {Kleiser}, {Van Dyk}, {Silverman}, {Graham}, {Kelly}, {Filippenko}, \& {Kumar}}]{2016MNRAS.461.3057S}
{Shivvers}, I., {Zheng}, W.~K., {Mauerhan}, J., {et~al.} 2016, \bibinfo{title}{{SN 2015U: a rapidly evolving and luminous Type Ibn supernova},} \mnras, 461, 3057, \dodoi{10.1093/mnras/stw1528}

% type= article
\bibitem[{I. {Shivvers} {et~al.}(2017){Shivvers}, {Zheng}, {Van Dyk}, {Mauerhan}, {Filippenko}, {Smith}, {Foley}, {Mazzali}, {Kamble}, {Kilpatrick}, {Margutti}, {Yuk}, {Graham}, {Kelly}, {Andrews}, {Matheson}, {Wood-Vasey}, {Ponder}, {Brown}, {Chevalier}, {Milisavljevic}, {Drout}, {Parrent}, {Soderberg}, {Ashall}, {Piascik}, \& {Prentice}}]{2017MNRAS.471.4381S}
{Shivvers}, I., {Zheng}, W., {Van Dyk}, S.~D., {et~al.} 2017, \bibinfo{title}{{The nearby Type Ibn supernova 2015G: signatures of asymmetry and progenitor constraints},} \mnras, 471, 4381, \dodoi{10.1093/mnras/stx1885}

% type= article
\bibitem[{R. {Siebenmorgen} {et~al.}(2014){Siebenmorgen}, {Voshchinnikov}, \& {Bagnulo}}]{2014A&A...561A..82S}
{Siebenmorgen}, R., {Voshchinnikov}, N.~V., \& {Bagnulo}, S. 2014, \bibinfo{title}{{Dust in the diffuse interstellar medium. Extinction, emission, linear and circular polarisation},} \aap, 561, A82, \dodoi{10.1051/0004-6361/201321716}

% type= article
\bibitem[{J.~F.~L. {Simmons} \& B.~G. {Stewart}(1985){Simmons} \& {Stewart}}]{1985A&A...142..100S}
{Simmons}, J.~F.~L., \& {Stewart}, B.~G. 1985, \bibinfo{title}{{Point and interval estimation of the true unbiased degree of linear polarization in the presence of low signal-to-noise ratios},} \aap, 142, 100

% type= article
\bibitem[{N. {Smith}(2014){Smith}}]{Smith2014}
{Smith}, N. 2014, \bibinfo{title}{{Mass Loss: Its Effect on the Evolution and Fate of High-Mass Stars},} \araa, 52, 487, \dodoi{10.1146/annurev-astro-081913-040025}

% type= incollection
\bibitem[{N. {Smith}(2017){Smith}}]{2017hsn..book..403S}
{Smith}, N. 2017, \bibinfo{title}{{Interacting Supernovae: Types IIn and Ibn},} in Handbook of Supernovae, ed. A.~W. {Alsabti} \& P.~{Murdin} (Springer International Publishing), 403, \dodoi{10.1007/978-3-319-21846-5_38}

% type= article
\bibitem[{N. {Smith} \& W.~D. {Arnett}(2014){Smith} \& {Arnett}}]{Smith_Arnett_2014}
{Smith}, N., \& {Arnett}, W.~D. 2014, \bibinfo{title}{{Preparing for an Explosion: Hydrodynamic Instabilities and Turbulence in Presupernovae},} \apj, 785, 82, \dodoi{10.1088/0004-637X/785/2/82}

% type= article
\bibitem[{N. {Soker}(2013){Soker}}]{Soker2013}
{Soker}, N. 2013, \bibinfo{title}{{A binary scenario for the pre-explosion outburst of the supernova 2010mc},} arXiv e-prints, arXiv:1302.5037, \dodoi{10.48550/arXiv.1302.5037}

% type= article
\bibitem[{S. {Srivastav} {et~al.}(2026){Srivastav}, {Smartt}, {Moore}, {Smith}, {Young}, {Fulton}, {Angus}, {Nicholl}, {Stevance}, {Chen}, \& et~al.}]{Srivastav2026}
{Srivastav}, S., {Smartt}, S.~J., {Moore}, T., {et~al.} 2026, \bibinfo{title}{{ATLAS100 ─ I. A volume-limited sample of supernovae and related transients within 100 Mpc},} \mnras, 549, stag1028, \dodoi{10.1093/mnras/stag1028}

% type= article
\bibitem[{J.~M. {Stone} {et~al.}(2020){Stone}, {Tomida}, {White}, \& {Felker}}]{2020ApJS..249....4S}
{Stone}, J.~M., {Tomida}, K., {White}, C.~J., \& {Felker}, K.~G. 2020, \bibinfo{title}{{The Athena++ Adaptive Mesh Refinement Framework: Design and Magnetohydrodynamic Solvers},} \apjs, 249, 4, \dodoi{10.3847/1538-4365/ab929b}

% type= article
\bibitem[{A. {Suzuki} {et~al.}(2019){Suzuki}, {Moriya}, \& {Takiwaki}}]{2019ApJ...887..249S}
{Suzuki}, A., {Moriya}, T.~J., \& {Takiwaki}, T. 2019, \bibinfo{title}{{Supernova Ejecta Interacting with a Circumstellar Disk. I. Two-dimensional Radiation-hydrodynamic Simulations},} \apj, 887, 249, \dodoi{10.3847/1538-4357/ab5a83}

% type= article
\bibitem[{A. {Suzuki} {et~al.}(2020){Suzuki}, {Moriya}, \& {Takiwaki}}]{Suzuki2020}
{Suzuki}, A., {Moriya}, T.~J., \& {Takiwaki}, T. 2020, \bibinfo{title}{{A Systematic Study on the Rise Time-Peak Luminosity Relation for Bright Optical Transients Powered by Wind Shock Breakout},} \apj, 899, 56, \dodoi{10.3847/1538-4357/aba0ba}

% type= article
\bibitem[{M. {Tanaka} {et~al.}(2017){Tanaka}, {Maeda}, {Mazzali}, {Kawabata}, \& {Nomoto}}]{Tanaka2017}
{Tanaka}, M., {Maeda}, K., {Mazzali}, P.~A., {Kawabata}, K.~S., \& {Nomoto}, K. 2017, \bibinfo{title}{{Three-dimensional Explosion Geometry of Stripped-envelope Core-collapse Supernovae. II. Modeling of Polarization},} \apj, 837, 105, \dodoi{10.3847/1538-4357/aa6035}

% type= article
\bibitem[{M. {Tanaka} {et~al.}(2009){Tanaka}, {Kawabata}, {Maeda}, {Iye}, {Hattori}, {Pian}, {Nomoto}, {Mazzali}, \& {Tominaga}}]{2009ApJ...699.1119T}
{Tanaka}, M., {Kawabata}, K.~S., {Maeda}, K., {et~al.} 2009, \bibinfo{title}{{Spectropolarimetry of the Unique Type Ib Supernova 2005bf: Larger Asymmetry Revealed by Later-Phase Data},} \apj, 699, 1119, \dodoi{10.1088/0004-637X/699/2/1119}

% type= inproceedings
\bibitem[{D. {Tody}(1986){Tody}}]{1986SPIE..627..733T}
{Tody}, D. 1986, \bibinfo{title}{{The IRAF Data Reduction and Analysis System},} in Society of Photo-Optical Instrumentation Engineers (SPIE) Conference Series, Vol. 627, Instrumentation in astronomy VI, ed. D.~L. {Crawford}, 733, \dodoi{10.1117/12.968154}

% type= inproceedings
\bibitem[{D. {Tody}(1993){Tody}}]{1993ASPC...52..173T}
{Tody}, D. 1993, \bibinfo{title}{{IRAF in the Nineties},} in Astronomical Society of the Pacific Conference Series, Vol.~52, Astronomical Data Analysis Software and Systems II, ed. R.~J. {Hanisch}, R.~J.~V. {Brissenden}, \& J.~{Barnes}, 173

% type= article
\bibitem[{D. {Tsuna} {et~al.}(2024){Tsuna}, {Wu}, {Fuller}, {Dong}, \& {Piro}}]{Tsuna2024}
{Tsuna}, D., {Wu}, S.~C., {Fuller}, J., {Dong}, Y., \& {Piro}, A.~L. 2024, \bibinfo{title}{{Merger Precursor: Year-long Transients Preceding Mergers of Low-mass Stripped Stars with Compact Objects},} The Open Journal of Astrophysics, 7, 82, \dodoi{10.33232/001c.123897}

% type= article
\bibitem[{P.~G. {van Dokkum}(2001){van Dokkum}}]{2001PASP..113.1420V}
{van Dokkum}, P.~G. 2001, \bibinfo{title}{{Cosmic-Ray Rejection by Laplacian Edge Detection},} \pasp, 113, 1420, \dodoi{10.1086/323894}

% type= article
\bibitem[{S.~S. {Vasylyev} {et~al.}(2023){Vasylyev}, {Yang}, {Filippenko}, {Patra}, {Brink}, {Wang}, {Chornock}, {Margutti}, {Gates}, {Burgasser}, {Karpoor}, {LeBaron}, {Softich}, {Theissen}, {Wiston}, \& {Zheng}}]{2023ApJ...955L..37V}
{Vasylyev}, S.~S., {Yang}, Y., {Filippenko}, A.~V., {et~al.} 2023, \bibinfo{title}{{Early Time Spectropolarimetry of the Aspherical Type II Supernova SN 2023ixf},} \apjl, 955, L37, \dodoi{10.3847/2041-8213/acf1a3}

% type= article
\bibitem[{S.~S. {Vasylyev} {et~al.}(2026){Vasylyev}, {Dessart}, {Yang}, {Filippenko}, {Patra}, {Brink}, {Wang}, {Chornock}, {Margutti}, {Gates}, {Burgasser}, {Sears}, {Karpoor}, {LeBaron}, {Softich}, {Theissen}, {Wiston}, \& {Zheng}}]{2026ApJ..1000...18V}
{Vasylyev}, S.~S., {Dessart}, L., {Yang}, Y., {et~al.} 2026, \bibinfo{title}{{Spectropolarimetric Evolution of SN 2023ixf: An Asymmetric Explosion in a Confined Aspherical Circumstellar Medium},} \apj, 1000, 18, \dodoi{10.3847/1538-4357/ae323f}

% type= article
\bibitem[{M. {Vetter} {et~al.}(2024){Vetter}, {R{\"o}pke}, {Schneider}, {Pakmor}, {Ohlmann}, {Lau}, \& {Andrassy}}]{2024A&A...691A.244V}
{Vetter}, M., {R{\"o}pke}, F.~K., {Schneider}, F. R.~N., {et~al.} 2024, \bibinfo{title}{{From spherical stars to disk-like structures: 3D common-envelope evolution of massive binaries beyond inspiral},} \aap, 691, A244, \dodoi{10.1051/0004-6361/202451579}

% type= article
\bibitem[{P. {Virtanen} {et~al.}(2020){Virtanen}, {Gommers}, {Oliphant}, {Haberland}, {Reddy}, {Cournapeau}, {Burovski}, {Peterson}, {Weckesser}, {Bright}, {van der Walt}, {Brett}, {Wilson}, {Millman}, {Mayorov}, {Nelson}, {Jones}, {Kern}, {Larson}, {Carey}, {Polat}, {Feng}, {Moore}, {VanderPlas}, {Laxalde}, {Perktold}, {Cimrman}, {Henriksen}, {Quintero}, {Harris}, {Archibald}, {Ribeiro}, {Pedregosa}, {van Mulbregt}, \& {SciPy 1. 0 Contributors}}]{2020NaMet..17..261V}
{Virtanen}, P., {Gommers}, R., {Oliphant}, T.~E., {et~al.} 2020, \bibinfo{title}{{SciPy 1.0: fundamental algorithms for scientific computing in Python},} Nature Methods, 17, 261, \dodoi{10.1038/s41592-019-0686-2}

% type= article
\bibitem[{L. {Wang} {et~al.}(2001){Wang}, {Howell}, {H{\"o}flich}, \& {Wheeler}}]{2001ApJ...550.1030W}
{Wang}, L., {Howell}, D.~A., {H{\"o}flich}, P., \& {Wheeler}, J.~C. 2001, \bibinfo{title}{{Bipolar Supernova Explosions},} \apj, 550, 1030, \dodoi{10.1086/319822}

% type= article
\bibitem[{L. {Wang} \& J.~C. {Wheeler}(2008){Wang} \& {Wheeler}}]{2008ARA&A..46..433W}
{Wang}, L., \& {Wheeler}, J.~C. 2008, \bibinfo{title}{{Spectropolarimetry of supernovae.},} \araa, 46, 433, \dodoi{10.1146/annurev.astro.46.060407.145139}

% type= article
\bibitem[{L. {Wang} {et~al.}(1997){Wang}, {Wheeler}, \& {H{\"o}flich}}]{1997ApJ...476L..27W}
{Wang}, L., {Wheeler}, J.~C., \& {H{\"o}flich}, P. 1997, \bibinfo{title}{{Polarimetry of the Type IA Supernova SN 1996X},} \apjl, 476, L27, \dodoi{10.1086/310495}

% type= article
\bibitem[{L. {Wang} {et~al.}(2003){Wang}, {Baade}, {H{\"o}flich}, {Khokhlov}, {Wheeler}, {Kasen}, {Nugent}, {Perlmutter}, {Fransson}, \& {Lundqvist}}]{2003ApJ...591.1110W}
{Wang}, L., {Baade}, D., {H{\"o}flich}, P., {et~al.} 2003, \bibinfo{title}{{Spectropolarimetry of SN 2001el in NGC 1448: Asphericity of a Normal Type Ia Supernova},} \apj, 591, 1110, \dodoi{10.1086/375444}

% type= article
\bibitem[{B. {Warwick} {et~al.}(2025){Warwick}, {Lyman}, {Pursiainen}, {Coppejans}, {Galbany}, {Jones}, {Killestein}, {Kumar}, {Oates}, {Ackley}, {Anderson}, {Aryan}, {Breton}, {Chen}, {Clark}, {Dhillon}, {Dyer}, {Gal-Yam}, {Galloway}, {Guti{\'e}rrez}, {Gromadzki}, {Inserra}, {Jim{\'e}nez-Ibarra}, {Kelsey}, {Kotak}, {Kravtsov}, {Kuncarayakti}, {Magee}, {Matilainen}, {Mattila}, {M{\"u}ller-Bravo}, {Nicholl}, {Noysena}, {Nuttall}, {O'Brien}, {O'Neill}, {Pall{\'e}}, {Pessi}, {Petrushevska}, {Pignata}, {Pollacco}, {Ragosta}, {Ramsay}, {Sahu}, {Sahu}, {Singh}, {Sollerman}, {Stanway}, {Starling}, {Steeghs}, {Teja}, \& {Ulaczyk}}]{Warwick2025}
{Warwick}, B., {Lyman}, J., {Pursiainen}, M., {et~al.} 2025, \bibinfo{title}{{SN 2023tsz: a helium-interaction-driven supernova in a very low-mass galaxy},} \mnras, 536, 3588, \dodoi{10.1093/mnras/stae2784}

% type= article
\bibitem[{X. {Wen} {et~al.}(2023){Wen}, {Gao}, {Ai}, {Liu}, {Zhu}, \& {Lei}}]{Wen2023}
{Wen}, X., {Gao}, H., {Ai}, S., {et~al.} 2023, \bibinfo{title}{{Polarization Signature of Companion-fed Supernovae Arising from BH-NS/BH Progenitor Systems},} \apj, 955, 9, \dodoi{10.3847/1538-4357/acef11}

% type= article
\bibitem[{S.~E. {Woosley}(2017){Woosley}}]{Woosley2017}
{Woosley}, S.~E. 2017, \bibinfo{title}{{Pulsational Pair-instability Supernovae},} \apj, 836, 244, \dodoi{10.3847/1538-4357/836/2/244}

% type= article
\bibitem[{S.~E. {Woosley} \& A. {Heger}(2015){Woosley} \& {Heger}}]{Woosley2015}
{Woosley}, S.~E., \& {Heger}, A. 2015, \bibinfo{title}{{The Remarkable Deaths of 9-11 Solar Mass Stars},} \apj, 810, 34, \dodoi{10.1088/0004-637X/810/1/34}

% type= article
\bibitem[{Y. {Yang} {et~al.}(2020{\natexlab{a}}){Yang}, {Hoeflich}, {Baade}, {Maund}, {Wang}, {Brown}, {Stevance}, {Arcavi}, {Burke}, {Cikota}, {Clocchiatti}, {Gal-Yam}, {Graham}, {Hiramatsu}, {Hosseinzadeh}, {Howell}, {Jha}, {McCully}, {Patat}, {Sand}, {Schulze}, {Spyromilio}, {Valenti}, {Vink{\'o}}, {Wang}, {Wheeler}, {Yaron}, \& {Zhang}}]{2020ApJ...902...46Y}
{Yang}, Y., {Hoeflich}, P., {Baade}, D., {et~al.} 2020{\natexlab{a}}, \bibinfo{title}{{The Young and Nearby Normal Type Ia Supernova 2018gv: UV-optical Observations and the Earliest Spectropolarimetry},} \apj, 902, 46, \dodoi{10.3847/1538-4357/aba759}

% type= article
\bibitem[{Y. {Yang} {et~al.}(2020{\natexlab{b}}){Yang}, {Hoeflich}, {Baade}, {Maund}, {Wang}, {Brown}, {Stevance}, {Arcavi}, {Burke}, {Cikota}, \& et~al.}]{Yang2020}
{Yang}, Y., {Hoeflich}, P., {Baade}, D., {et~al.} 2020{\natexlab{b}}, \bibinfo{title}{{The Young and Nearby Normal Type Ia Supernova 2018gv: UV-optical Observations and the Earliest Spectropolarimetry},} \apj, 902, 46, \dodoi{10.3847/1538-4357/aba759}

% type= article
\bibitem[{Y. {Yang} {et~al.}(2023){Yang}, {Baade}, {Hoeflich}, {Wang}, {Cikota}, {Chen}, {Burke}, {Hiramatsu}, {Pellegrino}, {Howell}, {McCully}, {Valenti}, {Schulze}, {Gal-Yam}, {Wang}, {Filippenko}, {Maeda}, {Bulla}, {Yao}, {Maund}, {Patat}, {Spyromilio}, {Wheeler}, {Rau}, {Hu}, {Li}, {Andrews}, {Galbany}, {Sand}, {Shahbandeh}, {Hsiao}, \& {Wang}}]{2023MNRAS.519.1618Y}
{Yang}, Y., {Baade}, D., {Hoeflich}, P., {et~al.} 2023, \bibinfo{title}{{The interaction of supernova 2018evt with a substantial amount of circumstellar matter - An SN 1997cy-like event},} \mnras, 519, 1618, \dodoi{10.1093/mnras/stac3477}

% type= article
\bibitem[{Y. {Yang} {et~al.}(2025){Yang}, {Wen}, {Wang}, {Baade}, {Wheeler}, {Filippenko}, {Gal-Yam}, {Maund}, {Schulze}, {Wang}, {Ashall}, {Bulla}, {Cikota}, {Gao}, {Hoeflich}, {Li}, {Mishra}, {Patat}, {Patra}, {Vasylyev}, \& {Yan}}]{Yang2025}
{Yang}, Y., {Wen}, X., {Wang}, L., {et~al.} 2025, \bibinfo{title}{{An axisymmetric shock breakout indicated by prompt polarized emission from the type II supernova 2024ggi},} Science Advances, 11, 24.2925, \dodoi{10.1126/sciadv.adx2925}

% type= article
\bibitem[{S.-C. {Yoon} {et~al.}(2017){Yoon}, {Dessart}, \& {Clocchiatti}}]{Yoon2017}
{Yoon}, S.-C., {Dessart}, L., \& {Clocchiatti}, A. 2017, \bibinfo{title}{{Type Ib and IIb Supernova Progenitors in Interacting Binary Systems},} \apj, 840, 10, \dodoi{10.3847/1538-4357/aa6afe}

% type= article
\bibitem[{T. {Yoshida} {et~al.}(2016){Yoshida}, {Umeda}, {Maeda}, \& {Ishii}}]{Yoshida2016}
{Yoshida}, T., {Umeda}, H., {Maeda}, K., \& {Ishii}, T. 2016, \bibinfo{title}{{Mass ejection by pulsational pair instability in very massive stars and implications for luminous supernovae},} \mnras, 457, 351, \dodoi{10.1093/mnras/stv3002}

\end{thebibliography}
\bibliographystyle{aasjournalv7}



\end{document}